\documentclass[12pt]{article}

\usepackage{latexsym,amsmath,amssymb,theorem,epsfig}
\usepackage{cite}

\DeclareFontFamily{OT1}{rsfs10}{}
\DeclareFontShape{OT1}{rsfs10}{m}{n}{ <-> rsfs10 }{}
\DeclareMathAlphabet{\mathscript}{OT1}{rsfs10}{m}{n}

\numberwithin{equation}{section}

\newcommand{\ns}{\normalsize}

\newcommand{\tr}{\text{tr}}

\newcommand{\CC}{{\mathbf{C}}}

\def\a{\alpha}
\def\b{\beta}
\def\g{\gamma}

\def\z{\psi}
\def\k{\kappa}
\def\l{\lambda}
\def\m{\mu}

\def\r{\rho}
\def\s{\sigma}

\def\z{\zeta}

\def\D{\Delta}

\def\gsim{ \lower .75ex \hbox{$\sim$} \llap{\raise .27ex \hbox{$>$}} }
\def\lsim{ \lower .75ex \hbox{$\sim$} \llap{\raise .27ex \hbox{$<$}} }
\def\be{\begin{equation}}
\def\ee{\end{equation}}
\def\bea{\begin{eqnarray}}
\def\eea{\end{eqnarray}}

\def \td {\tilde}

\def \ha {{1 \ov 2}}

\def \sql {{\sqrt{\l}}\ }
\def \del{\partial}
\def \a {\alpha}

\def\ov{\over}
\def \ci {\cite}

\def \foot {\footnote}
\def \bi{\bibitem}
\def\la{\label}\def\foot{\footnote}\newcommand{\rf}[1]{(\ref{#1})}

\def \OO {{\cal  O}}\def \no {\nonumber}
\def \hOO {\hat \OO}
\def \R {{\rm q}}

\def \el  {{\rm r}}
\def \uu {{\rm u}}
\def \z {\zeta}
\def \N {{\cal N}}
\def \xx {{\rm x}}
\def \adss {$AdS_5 \times S^5$\ }

\def \rC  {{\rm C}}
\def  \lg  {\langle}
\def  \rg  {\rangle}
\def \C  {{\cal C}}  \def \CC {\C}
\def \FF {{\cal F}}

\def \CC  {{\cal C}}

\def \rF {{\rm F}}
\def \g {{\gamma} } \def \hD {\hat \D}

\def \edo {\end{document}}
\def \ru {{\rm u}}

\def \N {{\cal N}}
\def \VV   {{\rm V}} \def \XX {{\rm X}}\def \ep {\epsilon}
\theoremstyle{plain}

{\theorembodyfont{\rmfamily} }

\begin{document}

%%%%%%%%%%%%%%%%%%%%%%%%%%%%%%%%%%%%%%%%%%%%%%%%%%%%%%%%%%%%%%%%%%%%%%

\begin{titlepage}

\vspace{-5cm}

%\hfill
%Imperial-TP-AT-2010-05
%\vskip-1pt

\vspace{-5cm}

\title{
  \hfill{\small Imperial-TP-EB-2011-01  }  \\[1em]
   {\LARGE  Correlation function of null polygonal  Wilson loops\\
    with local operators}
\\[1em] }
\author{L.F. Alday$^1$,
   E.I. Buchbinder$^2$  and   A.A. Tseytlin$^{2,}$\footnote{Also at Lebedev
    Institute, Moscow.}
     \\[0.5em]
   {\ns ${\ }^1$Mathematical Institute, University of Oxford, Oxford OX1 3LB, U.K.}
 \\[0.0em]
   {\ns ${\ }^2$The Blackett Laboratory, Imperial College, London SW7 2AZ, U.K.}
}

\date{}

\maketitle

\begin{abstract}
{\small  We consider the  correlator $\lg W_n \OO \rg/\lg W_n \rg$
 of a  light-like   polygonal   Wilson loop  with $n$ cusps
with a local  operator  (like the dilaton 
%$ \sim \tr F^2_{mn} + ...$ 
or a
chiral primary scalar) in planar  $\N =4$ super Yang-Mills theory.
As a consequence of conformal symmetry, the 
main
 %Ark
 part of such correlator is a function $F$ of $3n-11$ conformal
 ratios.
% involving the  position of the operator
% and the  positions of the  cusps.
 The  first non-trivial case is $n=4$ when $F$  depends on just  one conformal ratio $\z$.  This makes  the  corresponding correlator  one of the simplest non-trivial  observables
 that one would  like to  compute   for generic  values of the  `t Hooft  coupling $\l$.
We compute  $F(\z,\lambda)$ at leading order in both the strong coupling regime
 (using  semiclassical \adss string theory)  and  the weak  coupling regime (using  perturbative
gauge theory). Some  results are also obtained for polygonal Wilson loops with more than four edges. Furthermore, we also discuss a  connection to the  relation
 between  a correlator of  local operators at null-separated  positions and
 cusped Wilson loop suggested in  arXiv:1007.3243.
}
\end{abstract}

\thispagestyle{empty}

\end{titlepage}

\section{Introduction}

Recent remarkable progress  in understanding the duality between
planar $\N$=4  super Yang-Mills theory and superstring theory in \adss based on integrability
  opens  up the  possibility of computing
various observables exactly in `t Hooft coupling $\l$ or in  string tension $ \sql \ov 2\pi$.
Most of the progress was achieved for the   scaling   dimensions $\Delta_i(\l)$ of primary operators $\OO_i$ which determine the   2-point functions  $\lg \OO(x^{(1)})  \OO(x^{(2)}) \rg $
 (for reviews see \ci{beis}). The next step is to understand  3-point functions
$\lg \OO_i (x^{(1)}) \OO_j(x^{(2)}) \OO_k (x^{(3)}) \rg $  which, in addition to $\D_i$,   are
determined by  non-trivial functions $C_{ijk}(\l)$.
Higher-point correlation functions, though in principle dictated  by the OPE, are much more complicated. For example, conformal invariance implies
that  a 4-point
correlator  $\lg \OO_1 (x^{(1)}) ... \OO_4(x^{(4)}) \rangle $  should, in general,  contain
  a  non-trivial function of  the two conformal  cross-ratios
$\ru_1 = { |x^{(12)}|^2  |x^{(34)}|^2  \ov |x^{(13)}|^2  |x^{(24)}|^2  }, \ \
\ru_2 = { |x^{(12)}|^2  |x^{(34)}|^2  \ov |x^{(14)}|^2  |x^{(23)}|^2  }$\  \
($x^{(ij)}_m \equiv x^{(i)}_m -  x^{(j)}_m$, $m=0,1,2,3$)  and $\l$.

Correlators of primary operators are natural observables in   CFT.  In addition, in a  gauge
theory,   one may consider also expectation values of
Wilson loops.  An important class of these, related  to gluon scattering amplitudes
(see \ci{AM1,koo,br}  and   \ci{refs} for reviews), are expectation values of Wilson loops in
the fundamental representation
 $\lg W_n \rg$  corresponding to   polygons
  built out of null lines with $n$ cusps (located at
  $\{x^{(i)}_m \}, \ i=1,..., n$ with  $|x^{(i,i+1)}|^2=0, \   x^{(n+1)}\equiv x^{(1)} $).
They were previously  studied at weak \ci{korkor} and at strong \ci{Martin,AM1} coupling.
Conformal invariance (broken by the presence of the cusps in a controllable fashion)
 implies \ci{Drum} that for $n=4,5$  these expectation values are
fixed functions of $x^{(i)}$  (depending on a few  $\l$-dependent coefficients, in particular,
 on  the cusp anomalous dimension \ci{polya,kor})
 while for $n > 5$ they should depend on $3n-15$  cross-ratios of the cusp coordinates.
The first  non-trivial example is $\lg W_6 \rg$  which
is   expressed in terms of a  function of $\l$ and  three  cross-ratios.
%A27
For recent
   progress in computing this function at weak and at strong  coupling see
    \ci{xec,Bubble,xecc}.

As suggested   in  \ci{ak}, there is a  close relation
 between certain  correlators
of local (BPS)  operators and expectation values of cusped  Wilson loops:
  a  correlator
$K_n=\lg \hOO (x^{(1)}) ... \hOO(x^{(n)}) \rangle $ of primary operators
(e.g., the  highest weight part of 20'  scalar)
located at  positions of the null  cusps is  proportional  to
the  expectation  value of
the  null polygon Wilson loop in the  adjoint representation (or to
 $\lg W_n \rg^2$ in
the planar approximation we will consider here).  More precisely,
$\lim_{|x^{(i,i+1)}| \to 0}   K_n/K_{n0} =\lg W_n \rg^2 $,
where $K_{n0} \sim \prod^{n}_{i=1} |x^{(i,i+1)}|^{-2} + ...$
is the most singular term in the tree-level ($\l=0$)  part of $K_n$.

In this paper we  study a new   observable that involves  both a local operator
 and a cusped
Wilson loop, i.e.  $\lg  W_n  \OO(a)  \rg$  ($a$ will denote the position of
 the local operator).\foot{For
BPS  (circular)  Wilson loops  and their generalizations
 such correlators were studied previously  in \ci{cor,za02,zp,gom,alt}, see also
 \ci{miwa}. The null polygon  loop is, in a sense, a natural generalization
 of a circular loop
as it is   ``locally-BPS''. Furthermore, this kind of polygons is closed under conformal transformations.}
One motivation is that  such correlators
  may lead to new  simple examples  where one may be able to
 interpolate from weak to strong coupling. In particular, in the first non-trivial case  $n=4$
 such correlator  happens to  be a function of  just
{\it one}  non-trivial conformal ratio formed from the coordinates of the cusps $x^{(i)}$  and the operator $a_m$
(for $n >4$ it will  be a function of   $3n-11$  conformal ratios).
For comparison,
 in the case  of a circular Wilson loop
  (which, in fact,
   may be viewed as an $n \to \infty$ limit of a regular null polygon)
   the  dependence of the correlator $\lg  W_\infty   \OO(a)  \rg$ on the
    location of the operator $a$
is completely fixed \ci{cor,gom,alt} by conformal invariance.
%AT
%(taking into the account an anomalous
%transformation  under inversions).
%This  should thus be a new simple observable to study.
Determining  such a function (both at weak and at strong coupling)
 should be easier than the function of  the two  conformal ratios in
the 4-point correlator    case or  the function of  the three  conformal ratios in the 6-cusp   Wilson
loop   case. We shall  demonstrate this below by explicitly computing the
leading contributions
to  $\lg  W_4  \OO(a)  \rg$ both at strong and at weak coupling (for $\OO$ 
%Ark
being  the dilaton or a chiral primary operator).

Another  motivation to  study such ``mixed''  correlators is that they
may shed  more light on  the %  \ci{ak}
 relation \ci{ak} between the  correlators of null-separated operators  and
 cusped Wilson loops
mentioned above. That  relation was verified  at weak coupling, but  checking
it explicitly at strong coupling
remains an important open problem. For example, one may start  with
a $(n+1)$ -point correlator and consider a limit in which only $n$ of
 the locations of the operators become null-separated  and attempt
 to relate this limit to
 $\lg  W_n  \OO(a)  \rg$  with  $a= x^{(n+1)}$.
 %\foot{The correlator
 % $\lg \OO_1 (x^{(1)}) ... \OO(x^{(n)}) \ \OO(a) \rangle $
 %can not,  of course,  be directly  proportional to  $(\lg  W_n  \OO(a)  \rg)^2$
% but  some relation  may still be possible.}

 More explicitly, since the  derivative of a correlator
 over the gauge coupling brings down
 a power of the  super YM  action  which is the same as the
 integrated dilaton operator,
 the relation $\lg \hOO (x^{(1)}) ... \hOO(x^{(n)}) \rg  \sim  \lg  W_n   \rg^2$
  implies that
%AT
  \be
   \lg \hOO (x^{(1)}) ... \hOO(x^{(n)}) \int d^4 a\  \OO_{dil}(a)
   \rg  \sim  2 \lg  W_n   \rg  \lg  W_n  \int d^4 a\  \OO_{dil}(a) \rg\ .  \la{ptr}\ee
   Assuming that
   the  integral  over $a$ can be omitted and, furthermore, the dilaton operator
   can be replaced  by a generic  local  operator one may  conjecture that
 $
  \lg \hOO (x^{(1)}) ... \hOO(x^{(n)}) \ \OO(a)
   \rg \ \sim \  \lg  W_n   \rg  \lg  W_n  \ \OO(a) \rg $,
   i.e. that
     \be  \lim_{|x^{(i)} - x^{(i+1)}| \to 0}
   { \lg \hOO (x^{(1)}) ... \hOO(x^{(n)}) \ \OO(a)
   \rg \ov   \lg \hOO (x^{(1)}) ... \hOO(x^{(n)})
   \rg} \ \ \sim \ \ {    \lg  W_n  \ \OO(a) \rg \ov \lg  W_n   \rg } \ . \la{121}
   \ee
 Finally, it would     be  very
  interesting to
understand what is a  counterpart of  $\lg  W_n  \OO(a)  \rg$
 on the ``T-dual''  \ci{AM1}
 scattering-amplitude side, e.g., if there is any relation to
 form factors given  by $ \lg  A(x^{(1)} ) ...A(x^{(n)} )  \OO(a)  \rg$ where $A$'s  stand  for local fields like vector
  potential, cf. \ci{amf}.\foot{The two are obviously related when
  the  operator is at zero momentum (i.e. integrated over $a$), but in general
  one expects a complicated non-local relation involving a sum over contributions of
  different types of operators.}

%want to  understand 3 point functions -- recent progress for HHL
%relation to semiclassics for correlators of  operators
%Korchemsky's ideas that in correlation functions certain cusped Wilson line
%can be replaced with the twist-2 operator (whatever this means).\\
%In general, it is important to compute OPE coefficients in the expansion of Wilson loops.\\
%In the case of the circular Wilson loop, its correlation function with local
%operator is fixed by symmetries at any given coupling. In the case of the rectangular
%Wilson loop it is fixed up to a function of one variable. This is more interesting than circle.
%But it is simple enough to hope that it could be found to all loops.}

\

Let us briefly review the  contents  of this  paper.
 In Section 2, we  shall use general  symmetry considerations
to determine the structure of the correlator  \rf{1.1} of a null  $n$-polygon Wilson loop and a
conformal primary operator. We shall explicitly discuss the case of $n=4$   where
the result will be expressed in terms of a
 function $F$ of  only one non-trivial   conformal ratio \rf{ges} depending on the locations of the operator and the cusps.
Taking the $|a|\to \infty$ limit then determines the corresponding OPE coefficient \ci{cor}.

In Section 3  we  explicitly compute the $n=4$ correlator
 at strong coupling  using semiclassical string theory methods \ci{cor,za02}, i.e.
evaluating  the vertex operator corresponding to $\OO$ on the string surface \ci{Martin,AM1}
ending on the null quadrangle.
We shall explicitly determine the strong-coupling form of the function $F$
for the two  cases:  when $\OO$ is the dilaton
  or is  a chiral  primary operator.
We shall also discuss the generalization to
the case of non-zero R-charge or angular momentum in $S^5$.

In Section 4 we note that  since the string world surface ending on a null quadrangle is related  \ci{KRTT}
(by a euclidean continuation and conformal transformation) to the  surface describing
folded spinning  string \ci{gkp}  in the  infinite spin limit \ci{ft},
 the correlator computed in Section 3  may  be
related to  the strong-coupling limit of 3-point correlator of two  infinite spin twist-2 operators and a
dilaton operator.
%AT
The latter correlator
 may be   computed \ci{bt1,rt} using similar  semiclassical methods \ci{zar,cos}.
We point out that while  the  integrands in the two  expressions are indeed the same,
the ranges of integration are  different. The two integrals, however,
are indeed proportional   for a special choice of the locations of the twist-2 operators.

In Section 5  we discuss   the computation of the correlator
$\lg  W_n  \OO(a)  \rg$ at strong coupling  for  higher  number of cusps $ n >4$.
Unfortunately, the  explicit form of the space-time solution  is not know in this case,
but using the approach of \ci{AM2}  we are able to compute the   correlator
numerically in   the limit when the dilaton is far away from the null polygon,
i.e. to find  the OPE  coefficient   corresponding to the dilaton
in the expansion of the Wilson loop in its size.

In Section 6  we turn to the evaluation of this correlator  at weak coupling, i.e.
in perturbative  gauge theory. We explicitly see that the
leading  term in $\lg  W_4  \OO(a)  \rg$
has a form consistent with the one expected on symmetry grounds
with the function $F(\z)=\l h_0  + O(\l^2)$, where $h_0$ is a constant.
We  consider a  generalization to $n >4$ and compute the
OPE coefficient for the dilaton in the case of a regular null  polygon for arbitrary $n$.
We also compute the leading order $\l$ term in $F$ in  the case 
of the regular null  polygon with $n=6$.

In Section 7 we summarize  our results and mention some open questions.
In Appendix A  we discuss the  general structure of the correlator 
$\lg  W_4  \OO(a)  \rg$. 
%Ark
%In  
% Appendix B  we describe conformal transformations which relate
% four-cusped Wilson loop  surfaces  to the  string surface ending
%on a single null cusp and use them  to provide an alternative way
%to determine the   conformal ratio  \rf{ges} entering the  $n=4$ correlator.
In Appendix B we consider  some analytic results which can be obtained
for $n >4$ even number of cusps in the limit when the dimension
of the local operator is very large.

%%%%%%%%%%%%%%%%%%%%%%%%%%%%%%%%%%%%%%%%%%%%%%%%%%%%%%%%%%%%%%%%%%%%%%%%%%%%%%%%%%%%

\section{Structure of correlation function of cusped Wilson loop and a
  local operator }

Below we will consider the correlation function
\be
{\cal C}(W_n, {a})=\frac{\langle W_n {\cal O} ({a})\rangle}{\langle W_n\rangle}\,,
\label{1.1}
\ee
where $W_n$ is a polygonal Wilson loop made out of $n$ null lines  (see Figure 1) and ${\cal O}$ is a local scalar operator inserted at a generic  point
${a}=\{a_m\}=(a_0, a_1, a_2, a_3)$.
While the  expectation value $\langle W_n\rangle$
of such   Wilson loops is known to have UV divergences due to the presence of the
cusps \ci{polya,kor,korkor}  (enhanced in the null case)
we will see that
the ratio~\eqref{1.1} is finite, i.e. does not require a regularization.
%\foot{If
% none of the lines of the polygons  extends to infinity,  eq.~\eqref{1.1}
%is also  IR finite. -- wrong -- finite also if 2 cusps at infinity }

%
\begin{figure}[ht]
\centering
\includegraphics[width=45mm]{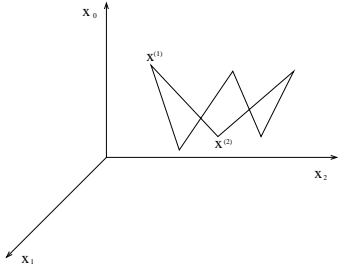}
\caption{\small Cusped polygonal Wilson loop, in this figure, with six edges. Consecutive cusps, for instance $x^{(1)}$ and $x^{(2)}$, are null separated.}
\label{fig1}
\end{figure}
\subsection{General considerations}

As follows from conformal symmetry, the non trivial part of
$\langle W_n\rangle$  depends only on the conformally invariant ratios constructed using
 the coordinates of the cusps~\cite{Drum}. The number of such conformal ratios for $n >5$  is
$4n-n -15=3n-15$. Here $4n$ stands for the total
 number of coordinates, $n$ is the number of
null  conditions  on the polygon lines
and $15$ is the dimension of the conformal group.\footnote{In~\cite{Drum}  this counting
was found using anomalous  Ward identities in the framework of
perturbative gauge theory.  In \cite{Bubble} the
 same result was found at strong coupling   by counting
the number of moduli of the  Hitchin's equation with certain boundary conditions
determining the corresponding minimal surface in $AdS_5$.} 
Furthermore, we expect \eqref{1.1} to be finite, since divergences from the numerator will be canceled by divergences from the denominator.

The number $3n-15$  of independent conformal ratios is exactly the same as the one that would  appear
in a correlator of $n\geq 4$ primary operators
%(e.g., scalar primaries as in~\cite{ak})
located at   the corners of a null polygon.
In general,  the structure of  $n$-point correlator  $\lg \OO(x^{(1)}) ... \OO(x^{(n)})\rg$
is fixed by conformal symmetry up to a function of conformal ratios.
The number  of these conformal ratios
is always given by $c_n=4n -\gamma_n$, where $4n$ is the total number of coordinates and
$\gamma_n$
is the number of generators of the conformal group broken by the precense of the local operators. For $n=2$, $3$, $4$
we have
$\g_2=8$, $\g_3=12$ and $\g_4=14$
so that    $c_2=0, \ c_3=0, \ c_4=2$.\foot{This  agrees with the familiar  count
of  the conformal ratios
$\uu^{(s)}= \prod^{n}_{i < j}  |x^{(i)}_m -x^{(j)}_m|^{2 \nu^{(s)}_{ij}}$.
Here $\nu^{(s)}_{ij}$ is a  basis in the space of
 symmetric  $n \times  n $  matrices $\nu_{ij}$. Scaling  invariance
and inversion symmetry
% (under which  $ |x -x'|^2 \to  {1 \ov |x|^2} {1 \ov |x'|^2} |x -x'|^2$)
imply that one should have $\nu_{ii}=0, \
\sum_{j=1}^{n} \nu_{ij} =0$ for  all $i=1, ..., n+1$.
This  leaves  $ \ha n (n+1) - n - n = \ha n (n-3)$  parameters
%A27
 \ci{syma}
but for $n >6$ not all of the corresponding  conformal ratios  are  functionally independent
%A27
(there are additional Gram determinant constraints).}
  A random  configuration of $n>4$ points breaks the conformal
group completely, i.e.   $\gamma_n=15$ and thus for $n >4$ we have
 $c_n= 4n -15$.\footnote{This
 can be seen, for example, as follows. For $n\geq 2$,
we can fix translations
and special conformal transformations by putting one point at the origin and one at infinity.
This configuration of two points
preserves dilatations and rotations which gives $7$ parameters implying  that
the number of the broken generators for $n=2$ is $8$. If we add one more point at some
arbitrary finite position we break dilatations and certain rotations.
What survives is the subgroup of the Lorentz group which preserves one vector. This
subgroup is 3-dimensional so that  $\gamma_3=12$.
If we add the fourth point the surviving subgroup has to preserve two vectors and, hence, is
one-dimensional, $\g_4=14$.
If we add one more point  all the conformal group becomes broken.}
If the operators are located at the corners of a  null polygon we have to impose $n$
additional constraints which gives $d_n=3n -\gamma_n$ for the number of conformal ratios,
i.e. $d_4=-2$,  $d_5 =0$ and thus  $d_n= 3n-15$ for $n  > 4$.

%However,   when $n<6$ points form a null polygon we do not have any conformal ratios left and, %hence, the counting
%$3n-15$ is correct for any $n$.

Adding an  operator ${\cal O}$ in~\eqref{1.1} at a generic point
brings in  4  parameters  so that $ {\cal C}(W_n, {a}) $ with $n \geq 4$
 should   be a non-trivial function
 %$F(\z_1,...,\z_{3n-11})$
 of $3n-11$
conformally invariant combinations $\z_k$
 constructed out of
the  coordinates $x_m^{(i)}$ of the $n$ cusps and the point $a_m$.\foot{We shall use
the notation $\z_k$  to distinguish these conformal ratios from standard cross-ratios
$\ru_k$ which appear in correlators of local operators at generic points.}
This is, of course, the
same as the number of conformal ratios  parametrising a correlator  of $n+1$
operators with only $n$ points being null-separated,
\be \la{redd}
c_{n+1} - n =  4(n+1) -15 - n =3n-11   \ . \ee
%One can view the correlator~\eqref{1.1} as a function not just of $a$ but also
%of the coordinates of the cusps. Then
%In addition, the  correlator \rf{1.1} involving the  primary operator $\OO$ should
%have a  prefactor consistent  with the scaling dimension $\D$  of the operator
%and expected OPE.
%AT18
Like for  correlators of primary   operators or  Wilson loops, the general  structure 
of ${\cal C} (W_n, a)$ in \rf{1.1}    should be  determined by  conformal invariance. We shall 
assume that in contrast to $\langle W_n\rangle$, which contains  UV divergences, the   correlator \rf{1.1}   should be UV finite (up to a possible renormalization of the
  operator $\OO$). As we shall argue below, in this  case  the 
   conformal invariance together with the expected OPE
property fixes ${\cal C} (W_n, a)$ up to a single function $F$
depending on $3n-11$ conformal ratios $\zeta_k$. 

In general, ${\cal C} (W_n, a)$ should be a function  of  $n$  distances 
$|a- x^{(k)}|$  and $\ha n (n-3)$  non-zero  ``diagonals'' of the null polygon 
$|x^{(i)}- x^{(j)}|$, $i \not= j\pm 1$.\foot{We shall  use the notation:
$|x-x'|^2= (x_m-x_m')^2 = -   (x_0-x_0')^2 +   (x_1-x_1')^2  +
 (x_2-x_2')^2   +(x_3-x_3')^2 $.} 
It should also transform  like  the operator 
$\OO(a)$  with dimension $\D$ 
under (i) dilatations and (ii) inversions, i.e. 
(i)  ${\cal C}\to   h^{-\D}{\cal C}$
under  $ x^{(i)} \to h  x^{(i)}, \ a\to h a$, and 
(ii) ${\cal C}\to   |a|^{2\D}{\cal C}$ under  $a_m \to \frac{a_m}{|x|^2}, \ 
x^{(i)}_m  \to   \frac{x^{(i)}_m}{ | x^{(i)}  |^2}$.\foot{Since special
conformal transformations are generated by translations and
inversions, it is enough to consider only the transformation under the inversions.
Note that under the inversions $ |x-x'|^2 \to \frac{| x  - x' |^2}{|x|^2 |x'|^2}$.}
The large $|a|$ behavior of ${\cal C}$  can be  fixed 
by consistency  with the expected  OPE expansion: for small Wilson
loop one may represent it
 in terms of a sum of local operators  \ci{shif,cor}
\be
{W_n \ov \langle W_n \rangle} = 1 + \sum_{k} c_k \  \el^{\Delta_k}\ \OO_k(0)  + ... \ , \la{22}
\ee
where
$\el$ is the characteristic  size of a loop,
$\OO_k$ are conformal primary operators with dimensions $\Delta_k$, and
dots stand for contributions of their conformal descendants.\foot{For example, in pure YM
  theory \ci{shif}:
$\lg W(C) \rg_{C\to 0} =1 + c_0 \el^4 \lg \OO_{dil.0} \rg +...$,
where $\OO_{dil.0} \sim  {1 \ov N} \tr F^2_{mn}$,
  $\el^4 $ stands for the square of the area of a disc bounded by $C$
and $c_0 =a_1 g^2 + a_2 g^4 + ...$. }
Taking  the position $a$   of the  operator ${\cal O}$ to be far away from the
null  polygon one should  then  get
\be
{\langle W_n {\cal O} ({a})\rangle  \ov \langle W_n \rangle}\Big|_{|a|\to \infty}
\  \sim \   \langle {\cal O}^{\dagger}(0) {\cal O} ({a})\rangle\  \sim
  \frac{1}{|{a}|^{2\Delta}}\,,
\label{1.3}
\ee
where ${\cal O}^{\dagger}$ conjugate to ${\cal O} $ is among the
operators present in \rf{22}. Since  all distances  $|a-x^{(k)}|$ 
between the operator and the cusps 
should  appear on an equal footing this   suggests the following  ansatz 
\be
{\cal C}(W_n, a)=
\frac{\FF(a,x^{(i)})}{\prod_{k=1}^n
|a-x^{(k)}|^{{2\ov n}\D}}\,,
\label{e1}
\ee
where  $\FF$ is   finite in the$|a|\to \infty$ limit, 
i.e.  it may depend on  $|a- x^{(k)}|$  only through their ratios. 
The dependence of $\FF$ on  $|x^{(i)}- x^{(j)}|$  is  constrained 
by the transformations under  dilatations and inversions mentioned above  which implies that 
under these two transformations we should have 
\be\la{tra}
(i)\ 
\FF \to h^{\D} \FF  \ ,  \ \ \ \ \ \ \ \ \ \ \ (ii) \ \FF \to 
 (|x^{(1)}|\ldots |x^{(n)}|)^{-\frac{2 }{n}\D}  \FF \ . \ee
 % under the inversions.
These  conditions  are solved, e.g., 
by taking $\FF \sim \prod_{i< j-1}^n |x^{(i)}-x^{(j)}|^{\mu}$ with
$\mu= \frac{2 }{n (n-3)}\Delta$. In addition, $\FF$ may contain 
a factor $F$  depending only on conformal ratios $\z_k$ which is manifestly invariant under the 
dilatations and inversions. As we argue in Appendix A, 
this is, in fact, the  general  structure of $\cal C$, i.e. 
we are led  to the following  expression for \rf{1.1} 
\be
{\cal C}(W_n, a)= \frac{\prod^n_{i  < j-1} | x^{(i)}- x^{(j)}|^{ \frac{2 }
{n (n-3)}\Delta } }{\prod_{k=1}^n |a- x^{(k)}|^{{2\ov n}\D}}\ F(\z_1,
...,\zeta_{3n-11})\,.
\label{1.2}
\ee
In general,  $\D$ and $F$ in \rf{1.2}  may depend also on the coupling
$\l$, i.e. they may look different at weak and at strong coupling,  but the
general structure \rf{1.2} should be universal. 

The  same  structure  \rf{1.2}  follows also  from
the general form  of  the correlator of local operators if the relation \rf{121}
is assumed to be true. 
%A27
 As is  well know, conformal invariance implies that    a correlator of
$q$  primary operators $\OO_i(x^{(i)})$ of dimensions $\D_i$  at generic positions
should have the form
% we should expect
\bea \la{133}
&&\lg \OO_1 (x^{(1)}) ... \OO_q (x^{(q)}) \rg
=  T_q   \  \rF_q (\ru_1, ..., \ru_{c_q}) \ , \ \ \ \ \ \ \ \
T_q \equiv  \prod^q_{i < j} | x^{(i)}- x^{(j)}|^{ - \g_{ij} }\  , \\
&&
\g_{ij}= {2 \ov q-2} \Big( \D_i + \D_j - {1 \ov q-1}
 \sum^q_{k=1} \D_k\Big)  \ , \ \ \ \
\ \ \ c_4=2 \ , \ \ \ c_{q > 4} = 4q -15  \la{quq}
\ , \eea
where $\rF_q $ is a function of conformally-invariant  cross-ratios.
Considering $q=n+1$   with $n$ operators  being the same, $\OO_k = \hOO,
\ \D_k = \hD $  and  $\OO_{n+1} = \OO$, \  $ \D_{n+1} = \D$ we find
\be \la{kp}
{T_{n+1} } =  \prod^n_{i < j} | x^{(i)}- x^{(j)}|^{ - { 2 \ov  n-1} ( \hD - { 1 \ov n} \D)
 }
\prod^n_{k=1} | x^{(n+1)}- x^{(k)}|^{ - { 2 \ov  n}\D  }\ , \ \ \ \ \
{T_{n} } =  \prod^n_{i < j} | x^{(i)}- x^{(j)}|^{ - { 2 \ov  n-1}  \hD }\ ,
\ee
so that in the  ratio of the two correlators  in \rf{121}  we have
\be \la{rat}
{T_{n+1}\ov T_n  } =
{ \prod^n_{i < j} | x^{(i)}- x^{(j)}|^{  { 2 \ov  n(n-1)}  \D } \ov
 \prod^n_{k=1} | a- x^{(k)}|^{  { 2 \ov  n}  \D } }\ , \ \ \ \ \ \ \ \ \ \ \
a\equiv x^{(n+1)} \ .
\ee
To get a  non-trivial expression in the null-separation limit
$| x^{(i)}- x^{(i+1)}| \to 0$ we will need  to assume  that $n$ of such vanishing
factors in numerator of \rf{rat}  get cancelled  against
similar factors  in some  cross-ratios contained in $\rF_{n+1}/\rF_{n}$.
That will  change the powers  of  the
remaining $\ha n (n-1) -n= \ha n (n-3) $
non-zero factors $| x^{(i)}- x^{(j)}|$  in \rf{rat} and also  reduce the total number
of non-trivial conformal ratios (now denoted by $\z_r$)  by $n$ as in \rf{redd}.
The result will then have the same form as in \rf{1.2}.

Indeed, the  combination   one needs to multiply \rf{rat}  by to cancel
the vanishing  $| x^{(i)}- x^{(i+1)}|$ factors in the numerator  and to match
the prefactor in \rf{1.2} with $\mu_{ij}= {2 \D \ov n(n-3)}  $ is ($x^{(n+1)}\equiv x^{(1)}$)
\be
{ \prod^n_{i  < j-1} | x^{(i)}- x^{(j)}|^{  { 4 \ov  n(n-1)(n-3)}  \D } \ov
 \prod^n_{k=1} | x^{(k)}- x^{(k+1)}|^{  { 2 \ov  n(n-1)}  \D    } }\  .
 \la{mul}
 \ee
One can check that this expression is invariant under both dilatations and inversions  and  can thus be expressed in terms of cross-ratios.

%\foot{Such a prefactor would of course be expected also in a correlator  where %$W_n$ is replaced by
%$n$ same primary operators at the locations of the cusps.}
Let us note that one  could, in principle,  treat $x^{(i)}$
 and $a$  on a different  footing,   aiming at determining the dependence on $a_m$  for
 fixed   positions  of the cusps  $x^{(i)}$ viewed  as given parameters.
In this  case the  $|x^{(i)}-x^{(j)}|$ dependent factor in \rf{1.2}  could be
formally absorbed into the function $F$.

%AT118
One may wonder how the  rational powers  in \rf{1.2} may appear 
  in  a  weak-coupling  perturbation theory. The point is that one can 
  recover  integer powers  for  an  appropriate  $F$. 
 We  shall comment on this issue in
   Appendix A on the example of $n=5$ and $\D=4$.

\subsection{$n=4$ case}

Let us now look in detail at the first non-trivial example: $n=4$.\foot{The  case of $n=3$ is trivial as there is no solution for coordinates of a null triangle in real 4d Minkowski space.}
Here    the number of variables $\zeta_k$
is   $3\times 4 -11=1$, i.e.  $F$  should be
a function of a {\it single} variable $\zeta_1\equiv \zeta$.
For $n=5$ the number of conformal ratios is already 4.
This makes the correlation
function~\eqref{1.1} for $n=4$ a particularly interesting and simple case to study.

As it follows from the above discussion, this variable $\zeta$ can be viewed
as the unique conformal ratio which one can build out of  the coordinates
$x^{(i)}_m$ \ ($i=1, \dots, 4$) of 4
cusps  and the location    $a_m$  of the operator ${\cal O}$.
Assuming that the null quadrangle is  ordered as
$x^{(1)},x^{(2)},x^{(3)},x^{(4)}$ (i.e.
 $|x^{(1)}- x^{(2)}|^2=|x^{(2)}- x^{(3)}|^2= |x^{(3)}- x^{(4)}|^2
    =|x^{(4)}- x^{(1)}|^2= 0$) it is easy to see
that the unique non-trivial  conformally-invariant   combination of these 5 points is
%\foot{
%As is well known special conformal transformations
%are generated by inversions and translations.
%%The invariance under dilatations  is obvious while the   invariance under the  inversions
%$x_m \to { x_m\ov |x|^2}$
%follows from the fact that for any two points  $x_m, \ x'_m$ we have
%$|x-x' |^2 \to
%\frac{| x  - x' |^2}{|x|^2 |x'|^2}\,. $ As is well known, the  special conformal transformations are generated by the  inversions and translations.  }
%
\be
\zeta=  \frac{ |a - x^{(2)} |^2\ | a  - x^{(4)} |^2 \  | x^{(1)}  - x^{(3)} |^2 }
  { |a- x^{(1)} |^2\
 |  a- x^{(3)} |^2 \ | x^{(2)}  - x^{(4)} |^2 } \ .
\label{ges}
\ee
In this  case there is  also a unique choice for the
$x^{(i)}$-dependent factor in \rf{1.2}:
$  ( | x^{(1)}  - x^{(3)} |  | x^{(2)}  - x^{(4)} |)^{\Delta/2}$
that ensures the right dimensionality of the result.
We  conclude that   the correlation function~\eqref{1.1}
for $n=4$  should have  the following general form
\be
{\cal C}(W_4, a)=\frac{( |x^{(1)} -x^{(3)}|  |x^{(2)} -x^{(4)}|)^{\Delta/2}}  {\prod_{i=1}^4 |a -x^{(i)}|^{\Delta/2}}\ F(\zeta)\,,
\label{1.15}
\ee
where $\Delta$ is the dimension of the operator $\OO$ and  $\zeta$ is given
by~\eqref{ges}.

%Thus, it remains to determine the function $F$.
As discussed above, the   same  conclusion applies also  to a correlator of 4  equivalent null-separated
operators  and an extra operator $\OO$. Indeed, for $n=4$ it is easy to see that
\rf{rat} is to be multiplied, according to \rf{mul},  by
\be
{  (| x^{(1)}- x^{(3)}| | x^{(2)}- x^{(4)}| ) ^{    \D/3 } \ov
 \prod^4_{k=1} | x^{(k)}- x^{(k+1)}|^{  \D/6    }
 }
 \  ,
 \la{mula}
 \ee
which is a product of two cross-ratios in power $\D/6$.

It is
interesting  to note that depending on just {\it one} conformal ratio, the $n=4$
correlator  \rf{1.15}
is an ``intermediate'' case
between a 3-point function
 $\langle \OO(x^{(1)}) \OO(x^{(2)})  \OO(x^{(3)})  \rangle$
which is completely fixed by conformal invariance (up to a function of the coupling)
and a  generic 4-point function  $\langle \OO(x^{(1)})...\OO(x^{(4)})  \rangle$
which depends on two conformal ratios.

In the limit when $|a| \to \infty$ we get
\bea
&& {\cal C} (W_4, a)_{|a| \to \infty}
 =\frac{\rC}{|a|^{2 \Delta}}\ , \label{1.16} \\  %\ , \ \ \ \ \ \ \ \ \
&&  \rC  \equiv  ( |x^{(1)} -x^{(3)}|  |x^{(2)} -x^{(4)}|)^{\Delta/2}  \ F (\z_\infty)  \ , \ \ \ \ \ \ \ \
\z_\infty= \frac{   | x^{(1)}  - x^{(3)} |^2}  {  | x^{(2)}  - x^{(4)} |^2 }
\,,
\label{1.166}
\eea
where  $ \rC $ thus determines the corresponding OPE coefficient in \rf{22}.

Another special limit is when the  position of the operator approaches
the location of one of the cusps, e.g., $a \to x^{(1)}$.
Setting  $a_m = x_m^{(1)} + \epsilon \alpha_m$,  $\epsilon \to 0$,  and using that
the vectors  $x^{(1)}-x^{(2)}$ and  $x^{(1)}-x^{(4)}$  are null
we find from~\eqref{ges} that $\zeta$ is, generically, finite in this limit and is given by
\be
\zeta_{a\to x^{(1)}}
 =  \frac{4\alpha \cdot (x^{(1)}-x^{(2)}){\ \ }\alpha \cdot (x^{(1)}-x^{(4)})    }
{\alpha^2\  |x^{(2)}-x^{(4)}|^2}\,, \ \ \ \ \ \ \ \
a_m = x_m^{(1)} + \epsilon \alpha_m \ .
\label{1.17}
\ee
Similarly, the limit of the pre-factor in~\eqref{1.15}  is
\be
\prod_{i=1}^4 |a-x^{(i)}|^{\Delta/ 2} {}_{_{a\to x^{(1)}}}
\ \to  \ \
 4 \ep^\Delta\    \a\cdot (x^{(1)}-x^{(2)} ) \ \a\cdot (x^{(1)}-x^{(4)} )\ |x^{(1)}-x^{(3)}|^2 \,.
\label{1.18}
\ee
Thus
\be {\cal C}( W_4^{(reg)}, {a}){}_{_{a\to x^{(1)}}}  \  \sim  \
  { 1 \ov  |a- x^{(1)}|^\D}   \ . \la{1.88} \ee
%AT
Note that this is the same behavior that would  be expected
 if the  Wilson loop were replaced  by a product of  4
same-type  operators  (e.g.,  scalar operators as in \ci{ak})
 at the positions of the cusps:
 $  \langle W_4 {\cal O} ({a})\rangle \to \langle
  \hOO(x^{(1)})...  \hOO(x^{(4)})\   {\cal O} ({a})\rangle$.
 Then the limit $ a\to x^{(1)}$ would  be determined by the OPE,
 $ \hOO  (x^{(1)})\ {\cal O} ({a})\ \sim   { 1 \ov
 |a - x^{(1)} |^{\D}}   \hOO  (x^{(1)}) $.
%  assuming   that there is a non-trivial
 %coupling between the operator ${\cal O}_\D$ and the  two operators
%$\hat \OO$  that   reproduce the Wilson loop.

One may also  consider a limit  when $a$ does not approach
a cusp but becomes null-separated   from it, i.e. $|a- x^{(1)}| \to 0$.
In  this  case the correlator will be divergent, i.e.
having $|a- x^{(i)}| \not=0$ is important for finiteness.
%A27
This is
analogous  to the observation in \ci{ak} that  keeping  $|x^{(i)}
- x^{(i+1)}|$ finite  in the  correlator of local operator  effectively regularizes the
null cusp  divergences of the corresponding Wilson loop.

\

Below we will explicitly verify the  general form  \rf{1.2},\rf{1.15} of the
 correlator \rf{1.1} at leading orders  in the
strong-coupling  (section 3)  and  the  weak-coupling (section 6)
expansions and compute the corresponding function $F$.

%%%%%%%%%%%%%%%%%%%%%%%%%%%%%%%%%%%%%%%%%%%%%%%%%%%%%%%%%%%%%%%%%%%%%%%%%%%%%%%%%%%%%%%%%%%%%%%%%%%%%%%%%%%%%%%%%%%%%%%%%%%%%%%%%

\section{Correlation function of  4-cusp  Wilson loop \\
with a local operator   %of cusped Wilson loop with local  operators
at strong coupling}

%%%%%%%%%%%%%%%%%%%%%%%%%%%%%%%%%%%%%%%%%%%%%%%%%%%%%%%%%%%%%%%%%%%%%%%%%%%%%%%%%%%%%%%%%%%%%%%%%%%%%%%%%%%%%%%%%%%%%%%%%%
In this section
we will compute   \rf{1.1} for  $n=4$  corresponding
to the 4-cusp  Wilson loop at strong coupling.
The result will have the expected form ~\eqref{1.15} and we will
explicitly determine  the function  $F(\zeta)$.
We shall  always consider the planar limit of maximally supersymmetric
Yang-Mills theory  and
assume that the operator
$\OO$ is such that  for large `t Hooft coupling $\l$  its   dimension
 $\Delta$ is much smaller than $\sqrt{\lambda}$.\foot{If the
 operator would carry  charges that are of order $\sql$ at
  large  coupling  it would modify the minimal surface that
   determines the leading order of the semiclassical expansion.}
  In particular,
 $\cal O$  will be chosen as  the dilaton operator or the  chiral primary operator.
We shall follow the same  semiclassical string theory
approach that was used  in the case of the  circular Wilson loop in   \ci{cor,za02} (see also \ci{zar,cos,rt,bt2,alt}).

%To calculate~\eqref{1.1}
%we will use the  same  approach as, e.g.,  in  \ci{cor,za02,zar,cos,rt,bt2,alt}.
% which was recently used for computation
%of three- and four-point functions~\cite{Zarembo, Costa, RT, Hern1, Ryang, Geor, BT2, Russo,
%Bak, Bissi, AR, Hern2, AT, Ahn}.
In string-theory  description the local operator ${\cal O}({a})$ is represented by a marginal
%integrated
 vertex operator~\cite{pt}
\be
\VV({a})=\int d^2 \xi\ V[X(\xi); {a}]\,,
\label{2.1}
\ee
where %$\xi$  parametrizes  worldsheet of disc topology
 $X$ stands for the 2d  fields  that enter the $AdS_5 \times S^5$ superstring action.
 In general,  \eqref{1.1} is then given by
\be
{\cal C}(W_n, {a})= \frac{1}{\langle W_n \rangle} \int  [dX] \   \VV(a) \  e^{-I[X]}
%\int d^2 \xi\ V[X(\xi); {a}]
\,.
\label{2.2}
\ee
Here $I$ is the string  action  proportional to the tension $T= {\sqrt \l \ov 2 \pi}$
and the path integral is performed over the euclidean
world-sheets with topology of a disc (we consider only the planar approximation)
and the boundary conditions  set out  by the Wilson loop at $z=0$.
Considering the  limit  when $ \sql \gg 1$  and assuming
that the operator represented by $\VV$ is ``light'' \ci{rt} (i.e. the corresponding
scaling dimension and  charges  are much smaller than $ \sql$)
one concludes that
this  path integral is dominated by the same semiclassical  string surface
as in the  absence of $\VV$, i.e. as  in the case of $\langle W_n \rangle$.
The resulting leading-order  value of \rf{2.2} is then given by \rf{2.1}
evaluated on this classical solution, i.e.
\be
{\cal C}(W_n, a)_{_{\sql \gg 1}}  = \Big( \int d^2 \xi\ V[X(\xi); a] \Big)_{semicl.}\,.
\label{2.3}
\ee
%
%where the right hand side is now understood as evaluated on the classical solution
%described above. Note that we can use the two-dimensional conformal invariance
%to go to an arbitrary coordinate system on the worldsheet. Note that unlike in the case
%of three-point functions of closed string states,
%now the worldsheet is, in general, of topology of a disc.

%%%%%%%%%%%%%%%%%%%%%%%%%%%%%%%%%%%%%%%%%%%%%%%%%%%%%%%%%%%%%%%%%%%%%%%%%%%%%%%%%%%%%%%%%%%%%%%%%%%%%%%%%%%%%%%%%%%%%%%%%%%

\subsection{Correlation function with  dilaton operator}

%%%%%%%%%%%%%%%%%%%%%%%%%%%%%%%%%%%%%%%%%%%%%%%%%%%%%%%%%%%%%%%%%%%%%%%%%%%%%%%%%%%%%%%%%%%%%%%%%%%%%%%%%%%%%%%%%%%%%%%%%%%

One simple case is  when the local operator $\OO$ is the
 dilaton operator $\OO_{dil}\sim  \tr ( F^2_{mn} Z^j)+...$
%AT
(where we included also the R-charge $j$ dependence).
The corresponding  vertex operator has the form~\cite{rt}
\bea
&&\VV_{dil} ({a}) =c_{dil} \int d^2 \xi \ \Big[\frac{z}{z^2+ (x_m-a_m)^2}\Big]^\Delta\ \XX^j \
U_{dil} \ ,
%{\cal L}\,,
\label{2.4}\\
&&  \XX^j = \big(\cos \theta \ e^{i \varphi}\big)^j  \ , \ \ \ \ \  \ \ \ \ \Delta = 4 + j  \ , \la{ku}
\eea
where $j\ll \sql $  is an angular  momentum along $S^1$ in $S^5$.
% the power 4  indicates the dimension of the operator $\Delta=4$,
The operator $U_{dil}$ equals the $AdS_5 \times S^5$ Lagrangian %
\be
U_{dil} ={\cal L}={\cal L}_{AdS_5} +  {\cal L}_{S^5} +{\rm fermions} \ , \ \ \ \ \ \
 {\cal L}_{AdS_5}= \frac{1}{z^2}[ (\partial_{\alpha}z)^2 +  (\partial_{\alpha}x_m)^2] \ .
\label{2.5}
\ee
Furthermore,
$c_{dil}$ is the normalization coefficient given by~\cite{cor,zar,rt}\foot{$N$
 is the  rank of  gauge group here representing a factor of string coupling.
 Note that the  normalization of the operator $\VV$  is important in order to compute
 the correlation function \rf{2.3}  and this   normalization is currently known only
 for the BPS  operators \cite{cor,zar,rt}.}
\be
c_{dil}= \frac{\sql }{8 \pi N}\sqrt{(j+1)(j+2)(j+3)} \,.
\label{2.6}
\ee
%
%Note that the proper normalization of the operator is important in order to compute
% the correlation function \rf{2.3}
Below  we shall mostly consider the case of $j=0$  when
\be
j=0: \ \ \ \ \ \ \ \ \ \   \ \ \ \Delta=4 \ , \ \ \ \ \ \   c_{dil}= \frac{\sqrt 6 \ \sql }{8 \pi N} \ ,
\ee
and return to the case of $j \not=0$ at the end of this subsection.

%%%%%%%%%%%%%%%%%%%%%%%%%%%%%%%%%%%%%%%%%%%%%%%%%%%%%%%%%%%%%%%%%%%%%%%%%%%%%%%%%%%%%%%%%%%%%%%%%%%%%%%%%%%%%%%%%%%%%%

\subsubsection{Regular 4-cusp case}

%%%%%%%%%%%%%%%%%%%%%%%%%%%%%%%%%%%%%%%%%%%%%%%%%%%%%%%%%%%%%%%%%%%%%%%%%%%%%%%%%%%%%%%%%%%%%%%%%%%%%%%%%%%%%%%%%%

Let us start with the case when the Wilson loop  is
the regular (i.e. equal-sided) quadrangle with 4 cusps
(Figure 2a).
\begin{figure}[ht]
\centering
\includegraphics[width=75mm]{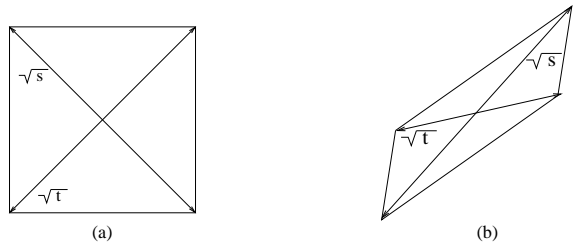}
\caption{\small $(x_1,x_2)$ plane
 projection of (a) regular and (b) irregular quadrangle  Wilson loop.}
\label{fig2}
\end{figure}
The classical euclidean world-sheet  surface  in $AdS_5$ ending on this Wilson loop was found in~\cite{AM1}
and is given by\foot{Here $(u,v)$
cover the full plane, but since infinity is not identified
the world sheet has topology of a disc.}
\bea
&& z=\frac{\el}{\cosh u\ \cosh v}\,, \ \ \ \qquad
x_0= \el\ \tanh u\ \tanh v\,, \nonumber\\
&&
x_1= \el\ \tanh u \,, \qquad x_2= \el\ \tanh v \,,\ \quad x_3=0\,; \ \ \quad u, v \in (-\infty, \infty)\,.
\label{1.4}
\eea
Here  $z$ is the radial direction of the Poincare patch of
 $AdS_5$ and $x_m=(x_0, x_1, x_2, x_3)$ are
the coordinates on the  boundary.
The parameter $\el$ corresponds to the overall scale  of the loop.
To simplify  later formulas
we will set $\el =1$ (it   is easy to restore $\el$ by simply replacing
$z \to \el^{-1} z  $, $x_m \to \el^{-1} x_m $).
The cusps correspond to
$(u, v) \to (\pm \infty, \pm \infty)$
and thus are located at
% (here $x=(x_0,x_1,x_2,x_3)$)
\bea
&& x^{(1)}=(1,\ 1,\ 1,\ 0)\,, \qquad\ \ \ \ \ \   x^{(2)}=(-1,\ 1, -1,\ 0)\,,\no  \\
&&
x^{(3)}=(1,-1,-1,\ 0)\,, \qquad\ \ \    x^{(4)}=(-1,-1,\ 1,\ 0)\,,
\label{2.16}
\eea
Substituting~\eqref{2.16} into~\eqref{ges} gives  the following explicit form of
the %corresponding
 conformal ratio $\z$ that is expected to appear in  the correlator
%AT
\bea
&& \zeta = \frac{(\ha  \R- a_0 - a_1 + a_2) (\ha \R- a_0 + a_1 - a_2)}
{ (\ha \R+ a_0 - a_1 - a_2) (\ha  \R+ a_0 + a_1 + a_2)} \ ,
\label{1.13} \\
&& \R\equiv 1-a_0^2+a_1^2+a_2^2 +a_3^2\,.
\label{1.14}
\eea
%A27
%An alternative way to arrive at eq.~\eqref{1.14}
%uses   the possibility to conformally map the 4-cusp surface
% to the single-cusp one of  \cite{Martin} (with 3 other cusps   moved to infinity).
%Ark
% see Appendix B.
Substituting the classical
solution~\eqref{1.4} into \eqref{2.4} we obtain\foot{Below in this
 section the expression
  for  a correlator will always stand for its leading $\sql \gg 1$ value.}
\be
{\cal C}_{dil}( W_4^{(reg)}, {a})=2 c_{dil} \int_{-\infty}^{\infty} d u d v \
\Big[ \frac{ (\cosh u\  \cosh v)^{-1} }
{\R -2 a_1 \tanh u -2 a_2 \tanh v + 2 a_0 \tanh u \tanh v} \Big]^4 \ ,
\label{2.7}
\ee
where $\R$ is given by~\eqref{1.14} and we used the fact that
on the solution~\eqref{1.4} one has $U_{dil}=2$ in \rf{2.15}
(note also that here
$\int d^2 \xi = \int dudv$).
 The integral is straightforward
to do by introducing the  variables $U=\tanh u,\  V=\tanh v$  and we get
\bea
&&{\cal C}_{dil}( W_4^{(reg)}, {a})=c_{dil}
\frac{16 a_1 a_2 - 8 \R a_0 - (\R^2 +4 a_0^2 -4 a_1^2 -4 a_2^2)\ \log \z}
{12(\R a_0 -2 a_1 a_2)^3}\,,
\label{2.8}
% \\
%&&\z= \frac{(\ha  R- a_0 - a_1 + a_2) (\ha R- a_0 + a_1 - a_2)}
%{ (\ha R+ a_0 - a_1 - a_2) (\ha  R+ a_0 + a_1 + a_2)} \ ,
%\label{2.9}
\eea
where we have used \eqref{1.13}.
The result is thus finite, in
 contrast to the area of the
4-cusp surface that requires a regularization \ci{AM1}. Let us note that
if we consider  the dilaton operator at
 zero momentum, i.e. integrate over the point $a$,
we will recover the divergent  area expression as
then the dilaton  vertex operator  with $\D=4$ in \rf{2.4}
will become proportional to the  string action
(the $[...]^\D$ factor  in \rf{2.4}  effectively provides a regularization for $\D
\not=0$). This is, of course, related to the fact that an  insertion of the zero-momentum
dilaton is equivalent to  taking a  derivative over  the string  tension  which brings down a
factor of the string action.

Let us now show that the result \eqref{2.8} is indeed consistent with eq.~\eqref{1.15}
for $\Delta=4$.
%For this we first note that
% \be L=-\log (\zeta-1)\,. \label{2.10} \ee
%
We observe that
\bea
&& \R^2+4 a_0^2 -4 a_1^2 -4a_2^2= \frac{1 +\zeta}{2}P_1 P_2\,,\ \  \ \
P_{1,2}\equiv  \R+2 a_0 \mp2 a_1 \mp2 a_2\,, %\ \  P_2= R+2 a_0 +2 a_1 +2 a_2\langle W_n {\cal O} ({a})\rangle\,.
 \no \\
&& \R a_0-2 a_1 a_2 =\frac{1-\zeta}{8}P_1 P_2\,.
\label{2.13}
\eea
If we substitute eqs.~\eqref{2.13} into~\eqref{2.8}
%and also multiply and divide by $\zeta$
 we get  (restoring   the dependence on the
scale parameter $\el$ in \rf{1.4})
\be
{\cal C}_{dil}( W_4^{(reg)}, {a})=\frac{64  \el^4 c_{dil} }{3} \ \frac{1}{P_1^2 P_2^2  } \ \frac{1}{(\zeta-1)^3}
[-2(\zeta-1)  +(\zeta +1) \log  \zeta]
 \,.
\label{2.14}
\ee
Finally, one can check  that
\be
 P_1^2 P_2^2= \zeta^{-1} \prod_{i=1}^4 |a- x^{(i)}|^2\,, \ \ \ \ \ \ \ \ \ \
 |x^{(1)}- x^{(3)}  |^2  |x^{(2)}-  x^{(4)}  |^2 = 64 \el^4 \ ,
\label{2.15}
\ee
where $x^{(i)}$ are the locations \rf{2.16} of the cusps in~\eqref{1.4}.
We conclude that the correlator
\rf{2.7} is given by  eq.~\eqref{1.15} with
\be
F(\zeta) =\frac{ c_{dil}}{3}\ \frac{\zeta}{(\zeta-1)^3}[-2(\zeta-1) +(\zeta +1) \log \zeta]\,.
\label{2.17}
\ee
In the limit $|a|\to \infty$ (see \rf{1.16})  we get $\z_\infty=1$  and thus
\be
{\cal C}_{dil}( W_4^{(reg)}, {a})_{_{|a|\to \infty} }
\  =\   \frac{32  c_{dil}\ \el^4 }{9 \ |a|^8}\,.
\label{2.18}
\ee
which determines
%From eqs.~\eqref{2.8}, \eqref{2.14} it is straightforward to find
 the OPE coefficient
of ${\cal O}_{dil}$ in the expansion \rf{22} of the Wilson loop
 $W_{4}^{(reg)}$.\foot{
%Taking the position $a_m$ of the operator  to be far away from
%the rectangular loop  along a space-like (e.g., $a_3$) direction we  get
The same expression can be obtained
by taking $|{a}|$  large directly in~\eqref{2.4} and  doing the
resulting simple integral   $\sim |a|^{-8} \int d^2 \xi \ z^4$.}

In the limit when $a$ approaches a  cusp ($a_m=x^{(1)}_m + \ep \alpha_m, \ \ \ep \to 0$)
 (see \rf{1.17},\rf{1.18})
we get
%Let us now comment on  what happens when the
%operator approaches one of  the cusps.
%Explicitly, we set $\D=4$ and
%\be
%a_0 = 1+ \ep  \a_0\,, \quad a_1 = 1+ \ep  \a_1\,, \quad
%a_2= 1+ \ep  \a_2\,, \quad a_3 = \ep  \a_3\,, \ \ \ \ \ep \to 0 \ .
%\label{2.28.1}\ee
%
%Then expanding~\eqref{2.8} for small $\ep$ yields
%
\bea
&&{\cal C}_{dil}( W_4^{(reg)}, {a})_{a\to x^{(1)}}
\  \to \
-\frac{2}{3 \ep^4}\  \frac{1}{[-3 \a_0^2+ \a_3^2 +(\a_1-\a_2)^2 +2 \a_0 (\a_1+\a_2)]^2 }
\nonumber\\
&&
\times \Big[1 - \frac{\a_3^2+ (\a_1+\a_2-\a_0)^2}{-3 \a_0^2+ \a_3^2 +(\a_1-\a_2)^2 +2
\a_0 (\a_1+\a_2)  }
\log \frac{\a_1^2 +\a_2^2 +\a_3^2-\a_0^2}{2(\a_0-\a_1)(\a_0-\a_2)}\Big]\,.
\label{2.28.2}
\eea
The  behavior $\epsilon^{-\Delta}=\epsilon^{-4}$ is in
agreement with the general expression \rf{1.88}.\foot{Note that
eq.~\eqref{2.28.2} gives the general behavior near the cusp. However, if we approach
it along a specific path we can have additonal singularities which
can all be found from~\eqref{2.28.2}.}

Let  us note also that one may consider a different limit
when $a$ does  not approach the cusp
$x^{(1)}$ but becomes null separated from it, i.e.
$| a- x^{(1)}| \to 0$.  In this   case  the correlator is logarithmically divergent:
${\cal C}_{dil}( W_4^{(reg)}, {a})
\sim \log | a- x^{(1)}|$. If $a$ becomes at the same time  null-separated   from the
two adjacent cusps (say, $x^{(1)},x^{(2)}$)   then $\z$ stays finite  and the correlator has a power divergence
from the prefactor:
${\cal C}_{dil}( W_4^{(reg)}, {a})
\sim  | a- x^{(1)}|^{-2} | a- x^{(2)}|^{-2}$.

%{\bf comment of factorization in this limit}

%%%%%%%%%%%%%%%%%%%%%%%%%%%%%%%%%%%%%%%%%%%%%%%%%%%%%%%%%%%%%%%%%%%%%%%%%%%%%%%%%%%%%%%%%%%%%%%%%%%%%%%%%%%%%%%%%%%%%%

\subsubsection{Irregular 4-cusp case}

%%%%%%%%%%%%%%%%%%%%%%%%%%%%%%%%%%%%%%%%%%%%%%%%%%%%%%%%%%%%%%%%%%%%%%%%%%%%%%%%%%%%%%%%%%%%%%%%%%%%%%%%%%%%%%%%%%%%%

The above calculation can be generalized
to the case of an irregular quadrangle, i.e.  the one  with  unequal diagonals
$s \neq t$ (Figure 2b). The  corresponding  solution can be found by applying a conformal
transformation
to \rf{1.4} ~\cite{AM1} 
%(see also Appendix B)
%
\bea
&&
z= {f(u,v)}\ {\cosh u\ \cosh v }\,, \qquad
x_0= {\sqrt{1+b^2}\ f(u,v)\  \tanh u\ \tanh v}\,,  \nonumber \\
&&
x_1 =f(u,v)\ { \tanh u}\ , \qquad
%{1+ b \tanh u\ \tanh v}\,, \quad
x_2 =f(u,v)\ { \tanh v} \,, \qquad x_3=0\,, \nonumber\\
&&
f(u,v)\equiv  \frac{\el}{1+ b \tanh u\ \tanh v}\,, \ \ \ \ \ \ \ \ \  |b|\leq 1 \ .
\label{2.19}
\eea
$b=0$ corresponds to the regular quadrangle case \rf{1.4}.
The cusps
are found by taking $(u, v) \to (\pm \infty, \pm \infty)$ and are located at
(cf. \eqref{2.16}; here we set   $\el=1$)
\bea
&&
x^{(1)}_m=(\frac{\sqrt{1+b^2}}{1+b}, {\ } \frac{1}{1+b}, {\ }\frac{1}{1+b}, {\ }0)\,, \quad
x^{(2)}_m=(-\frac{\sqrt{1+b^2}}{1-b}, {\ } \frac{1}{1-b}, {\ }\frac{-1}{1-b}, {\ }0)\,,
\nonumber\\
&&
x^{(3)}_m=(\frac{\sqrt{1+b^2}}{1+b}, {\ } \frac{-1}{1+b}, {\ }\frac{-1}{1+b}, {\ }0)\,, \quad
x^{(4)}_m=(-\frac{\sqrt{1+b^2}}{1-b}, {\ } \frac{-1}{1-b}, {\ }\frac{1}{1-b}, {\ }0)\,.
\label{2.20}
\eea
The Wilson loop is the quadrangle $x^{(1)},x^{(2)},x^{(3)},x^{(4)}$.\foot{It is easy to check that
the vectors connecting the 4 cusps
$x^{(12)}=x^{(1)}-x^{(2)}$, $x^{(23)}=x^{(2)}-x^{(3)}$, $x^{(34)}=
x^{(3)}-x^{(4)}$, $x^{(41)}=x^{(4)}-x^{(1)}$ are null.
The two non-trivial parameters   $s$ and $t$ are given by~\cite{AM1}
%AT
$
-(2 \pi)^2 s = 2 x^{(23)} \cdot  x^{(34)}= | x^{(2)} - x^{(4)} |^2=   \frac{8}{(1-b)^2}\,, \ \ \ \
- (2 \pi)^2 t = 2 x^{(12)}\cdot  x^{(23)}= | x^{(1)} - x^{(3)} |^2= \frac{8}{(1+b)^2}\,.
$
}
%
%where by $x^{(i)}x^{(j)}$ we denoted the null vector connecting the cusps located
%at  $x^{(i)}_m$ and $x^{(j)}_m$.}

Since  ${U_{dil}}$ in \rf{2.5} is invariant under $SO(2,4)$, its   value on this solution should be
the same as for $b=0$, i.e. $U_{dil}=2$. Substituting~\eqref{2.19}
into ~\eqref{2.4} gives
\be
{\cal C}_{dil}( W_4^{(irreg)}, {a})=2 c_{dil} \int_{-\infty}^{\infty} d u d v
\ \Big[ \frac{(\cosh u\ \cosh v)^{-1} }
{\R -2 a_1 \tanh u -2 a_2 \tanh v + 2 \tilde{a}_0 \tanh u \tanh v} \Big]^4\,,
\label{2.22}
\ee
where  $\tilde{a}_0$ is defined by
\be
\tilde{a}_0 = a_0 \sqrt{1+b^2} + \ha  b (\R-2)  \ ,
\label{2.23}
\ee
while $\R$ is again  given by~\eqref{1.14} (without the replacement  $a_0\to \tilde{a}_0$).
As \rf{2.7} and ~\eqref{2.22} are related by replacing $a_0\to \tilde{a}_0$
we get from \eqref{2.8},\eqref{1.13}
\bea
&&{\cal C}_{dil}( W_4^{(irreg)}, {a})=c_{dil}
\frac{16 a_1 a_2 - 8 \R \tilde{a}_0 - (\R^2 +4 \tilde{a}_0^2 -4 a_1^2 -4 a_2^2)\ \log \z}
{12(\R \tilde{a}_0 -2 a_1 a_2)^3}\,,\label{2.24} \\
&&
%%\LL= \log \z \ , \ \ \ \ \ \ \ \
\z= \frac{ (\ha \R-2 \tilde{a}_0 - a_1 + a_2) (\ha \R- \tilde{a}_0 + a_1 - a_2)}
 {(\ha \R+ \tilde{a}_0 - a_1 - a_2)(\ha \R+ \tilde{a}_0 + a_1 + a_2)} \,.
\label{2.26}
\eea
It is  straightforward to check, using   the locations of the cusps
in~\eqref{2.20}, that the argument of the logarithm in \rf{2.26}
is again  the conformally-invariant ratio  in ~\eqref{ges}.
Note that while $b$ may be interpreted as the   parameter of a conformal
transformation relating the  regular and the  irregular polygons,
 $a$ is kept fixed under
this transformation, so  that $\z$ in \rf{2.26}  now depends on $b$ compared to the one in
\rf{1.13},\rf{2.8}.

After  the
same steps as in the case of the  regular quadrangle we find that eq.~\eqref{2.24}
can indeed be written as~\eqref{1.15}, where $\Delta=4$, $x^{(i)}$'s are given
by~\eqref{2.20} and
\bea
&&  |x^{(1)}- x^{(3)}  |^2  |x^{(2)}-  x^{(4)}  |^2 =
 {64 \el^4 \ov (1-b^2)^2} \  \la{bbb},\\
&&F(\zeta) =
  \frac{ c_{dil}}{3}\ \frac{\zeta}{(\zeta-1)^3}[-2(\zeta-1) +(\zeta +1) \log \zeta]\,.
\label{2.27}
\eea
The function   $F(\z)$ is thus  the   same as in \rf{2.17}, as expected.
%The OPE coefficient for the dilaton operator in  $W_4^{(irreg)}$
%is found again  by taking $|a| \to \infty$ in \rf{2.24} giving
%It is straightforward to find the OPE coefficient corresponding to ${\cal O}^{\dagger}_{dil}$ in the
%expansion of $W_4^{(irreg)}$ in terms of local operators. For this we again take the operator
%far away from the loop along a space-like (for example, along the $a_3$-) direction. We find that
%
We find also that
\be
{\cal C}_{dil}( W_4^{(irreg)}, {a})_{_{|a|\to \infty}}
 \ =   \ { {\rm C}_4 \ov |a|^8 }  \ , \ \ \ \ \ \ \ \ \
  {\rm C}_4=\frac{4c_{dil}\ \el^4 }{3b^3 \ }
\Big[-2b + (1+b^2)\log \frac{1+b}{1-b}  \Big]\,,
\label{2.28}
\ee
which reduces to ~\eqref{2.18} in   the limit $b \to 0$.

Let us note  also
that starting with the general expression for the correlator \rf{1.15},\rf{2.17}
we may consider    the  case (obtained by a conformal  transformation from \rf{1.14}  \ci{ryang})
when  two  of the four  cusps (e.g., $x^{(3)},x^{(4)}$)  are sent to infinity.
Then \rf{1.15} with $\D=4$ takes the following form:
\be \la{ry}
{\cal C}_{dil}( W_4)_{x^{(3)},x^{(4)}\to \infty} = {1 \ov
| a- x^{(1)}|^{2} | a- x^{(2)}|^{2}} \   F(\td \z)\ ,
\ \ \ \ \ \ \ \ \ \td \z = {| a- x^{(2)}|^{2}\ov | a- x^{(3)}|^2} \ . \ee
%%%%%%%%%%%%%%%%%%%%%%%%%%%%%%%

%%%%%%%%%%%%%%%%%%%%%%%%%%%%%%%%%%%%%%%%%%%%%%%%%%%%%%%%%%%%%%%%%%%%%%%%%%%%%%%%%%%%%%%%%%%%%%%%%%%%%%%%%%%%%%%%%%%%%%%%%%%%%

%\subsection{Correlation function with the dilaton carrying angular momentum}

\subsubsection{Case of dilaton  with non-zero $S^5$ momentum }

%%%%%%%%%%%%%%%%%%%%%%%%%%%%%%%%%%%%%%%%%%%%%%%%%%%%%%%%%%%%%%%%%%%%%%%%%%%%%%%%%%%%%%%%%%%%%%%%%%%%%%%%%%%%%%%%%%%%%%%%%%%%%%%

The above  discussion
 can be generalized to the case when the dilaton operator \rf{2.4}  carries an angular
momentum $j$ along $S^1 \subset S^5$. Again,
we have to
evaluate~\eqref{2.4} on the solution~\eqref{1.4} or~\eqref{2.19}. Since these solutions do not
depend on the sphere coordinates the factor $\XX^j$ in \rf{2.4}  equals to unity so that
the correlation function is given by the same expressions as in~\eqref{2.7} or \eqref{2.22}
with the power 4 replaced with $\Delta=4+j$.
 In the case of the more general
solution~\eqref{2.19} we get  instead of \rf{2.22}
\be
{\cal C}_{dil}( W_4^{(irreg)}, {a})=2 c_{dil} \int_{-\infty}^{\infty} d u d v
\ \Big[ \frac{ (\cosh u \ \cosh  v)^{-1} }
{\R -2 a_1 \tanh u -2 a_2 \tanh v + 2 \tilde{a}_0 \tanh u \tanh v} \Big]^{\Delta}
\label{2.30.1}
\ee
where $\tilde{a}_0$ is defined in \rf{2.23}.
%
%To find the answer for the case of regular Wilson loop we simply set $b=0$.
Since  the answer should be of the form~\eqref{1.15},
 we are interested  in computing the  function  $F(\zeta)$ of one variable only, so
 we  may  set $a_1=a_2=a_3=0$.
Having computed
  $\bar F(a_0) = F(\z(a_0))$ we may    find $a_0=a_0 (\zeta)$ from~\eqref{2.26}
 and thus restore $F(\z)$.
%Since $F$ depends only on one variable this procedure is unambiguous.
Performing the integral~\eqref{2.30.1} we obtain
\be
{\cal C}_{dil}( W_4^{(irreg)}, a_0)=\frac{2 \pi c_{dil}}{(1-a_0^2)^{\Delta}} \
\Big(\frac{\Gamma[\frac{\Delta}{2}]}{\Gamma[\frac{\Delta+1}{2}]}\Big)^2\
{\ }_2F_1(\frac{1}{2}, \frac{\Delta}{2}, \frac{\Delta+1}{2}, \varrho^2)\,,
\label{2.31}
\ee
where${\ }_2F_1$ is the hypergeometric function and $\varrho$ is a function of $a_0$
 given by
\be
\varrho\equiv  \frac{2 \tilde{a}_0}{1-a_0^2}= \frac{  2 a_0 \sqrt{1+b^2} -  b (1+a_0^2)}{1-a_0^2} \ .
\label{2.34}
\ee
To extract   $F$ in \rf{1.15}
we have to multiply~\eqref{2.31} by the  factor
$$
|x^{(1)}-  x^{(3)} |^{-\Delta/2}|x^{(2)}-  x^{(4)} |^{-\Delta/2}
 \prod_{i=1}^4 |x_{m}^{(i)}-a_m|^{\Delta/2}\,.
$$
This gives
\be
F(\z(\varrho))= {2^{-{3 \ov 2}  \D +1}  \pi \  c_{dil}}\ \Big(
\frac{\Gamma[\frac{\Delta}{2}]}{\Gamma[\frac{\Delta+1}{2}]}\Big)^2  \
(1-\varrho^2)^{\Delta/2}{\ }_2F_1(\frac{1}{2}, \frac{\Delta}{2}, \frac{\Delta+1}{2}, \varrho^2)\,.
\label{2.33}
\ee
%
%Note that it is a function only of $\varrho$.
Finally, we can express  $\varrho$ in terms of $\zeta$ using~\eqref{2.26}:
\be
\varrho = \frac{ 1- \sqrt{\zeta}}{1+ \sqrt \zeta }\,.
\label{2.35}
\ee
%
%so that  $F(\zeta)$ is given by eq.~\eqref{2.33} with $\varrho=\varrho(\zeta)$ as
%in~\eqref{2.35}.
One can check that setting $j=0$, i.e.  $\Delta =4$,
gives  back  our earlier  expression~\eqref{2.27}.

%{\bf  Large $\Delta$ limit ?? Semiclassics of the above integral over $u,v$ -- exists ?
% }

%%%%%%%%%%%%%%%%%%%%%%%%%%%%%%%%%%%%%%%%%%%%%%%%%%%%%%%%%%%%%%%%%%%%%%%%%%%%%%%%%%%%%%%%%%%%%%%%%%%%%%%%%%%%%%%%%%%%%%%%%%

\subsubsection{Generalization to cusped Wilson loop with an  $S^5$  momentum }

%%%%%%%%%%%%%%%%%%%%%%%%%%%%%%%%%%%%%%%%%%%%%%%%%%%%%%%%%%%%%%%%%%%%%%%%%%%%%%%%%%%%%%%%%%%%%%%%%%%%%%%%%%%%%%%%%%%%%%%

One  can formally repeat  the above discussion in  the case
when the euclidean 4-cusp world surface \rf{1.4} is generalized to the presence of a
non-zero angular momentum in $S^5$ (we again set the scale $\el=1$)\footnote{This solution is  related by an analytic  continuation
and a  conformal transformation  \ci{KRTT} to the large spin limit of the folded $(S,J)$ spinning
string  \ci{ft}, see also  section 4.}
\bea
&& z=\frac{1}{\cosh k u\  \cosh v}\,, \qquad \ \ \ \
x_0= \tanh k u\  \tanh v\,, \nonumber\\
&&
x_1= \tanh k u \,, \quad x_2= \tanh v \,, \quad x_3=0\ ,\ \ \ \ \  \ u, v \in (-\infty, \infty)\ ,\no \\
&&\varphi=- i \ell u\,, \ \  \ \ \ \ \ \ \ \  k^2 = 1 + \ell^2  \ .
\label{2.36}
\eea
 This background
\rf{2.36} solves the string equations in conformal gauge.
Here $\varphi$ is an  angle of a big circle  in $S^5$ and $\ell$
may be interpreted as a density of the corresponding
 angular momentum. We assume  that
$u$ plays the role of a euclidean time; since  $v$ is non-compact,
the total angular momentum is formally infinite.
The  presence of  $k$  does not  influence  the positions of the 4
 cusps  \rf{2.16}.

The correlation function with the  dilaton operator is again  given by \rf{2.3},\rf{2.4}.
Since now $\varphi$ in \rf{ku}
is non-zero\foot{In \rf{ku}  $\theta=0$ as we consider rotation in a big circle of $S^5$.}
 there will be a non-trivial
dependence on the product of the dilaton  momentum
 $j$ and the  angular momentum density $\ell$ (cf. \rf{2.7},\rf{2.22},\rf{2.30.1})
\bea
&&{\cal C}_{dil}( W_{4, \ell}^{(reg)}, {a})  \label{2.40} \\  &&=
2 c_{dil} \int_{-\infty}^{\infty} d u d v  \ \Big[ \frac{ (\cosh k u\  \cosh v)^{-1} }
{\R -2 a_1 \tanh k u -2 a_2 \tanh v + 2 a_0 \tanh k u\ \tanh v} \Big]^{4+j}\  e^{ j \ell   u} \ .
\no
\eea
We used that in \rf{2.5} ${\cal L}_{AdS_5} =k^2 +1\ , \ \ \
{\cal L}_{S^5}= (\partial_{\alpha}\varphi)^2=-\ell^2$  so that again $U_{dil}= {\cal L}=2$.
The resulting integral  can be studied for generic $j$ (it is useful to
change variable  $u \to u'=k u $).
 For $j=0$   we get back  to
the same expression as~\eqref{2.7}  with an extra factor of $ k^{-1} = (1 + \ell^2)^{-1/2}$, i.e.
\be j=0: \ \ \ \ \ \ \ \ \ \ \ \ \  \ \ \ \ \
{\cal C}_{dil}( W_{4,\ell }^{(reg)}, {a})=\frac{1}{\sqrt{1+\ell^2}}\
{\cal C}_{dil}( W_{4}^{(reg)}, {a})\,.
\label{2.41}
\ee
%
%That is the function $F(\zeta)$ is this case is given by
%
%\be
%F(\zeta)=\frac{64 c_{dil}}{3 \sqrt{1+\ell^2}}\frac{ 1-\zeta}{(2-\zeta)^3}[4-2\zeta +\zeta \log (\zeta-1)]\,.
%\label{2.42}
%\ee
%
%Generalization to the case of irregular Wilson loop is also straightforward
%and will lead to an additional factor $1/(1-b^2)^2$ in~\eqref{2.42}.
%It  is also straightforward to d
It is also straightfoward to discuss a generalization to
irregular 4-cusp surface with $b\not=0$ (cf. \rf{2.27}).

%%%%%%%%%%%%%%%%%%%%%%%%%%%%%%%%%%%%%%%%%%%%%%%%%%%%%%%%%%%%%%%%%%%%%%%%%%%%%%%%%%%%%%%%%%%%%%%%%%%%%%%%%%%%%%%%%%%%%%%%%%%%%%%%%%%

\subsection{Correlation function with chiral primary operator}

%%%%%%%%%%%%%%%%%%%%%%%%%%%%%%%%%%%%%%%%%%%%%%%%%%%%%%%%%%%%%%%%%%%%%%%%%%%%%%%%%%%%%%%%%%%%%%%%%%%%%%%%%%%%%%%%%%%%%%%%%%%%%%%

Let us now  consider a similar computation
with the  chiral primary  operator $\OO_j = \tr Z^j$
 instead of the dilaton operator.
%${\cal O}_j$.
The  bosonic part of the corresponding  vertex operator
\ci{cor,zar,rt}   can be written in a  form  similar to  \rf{2.4}
\bea
&& \VV_j ({a})=c_j \int d^2 \xi\   \Big[\frac{z}{z^2+(x_m-a_m)^2}\Big]^\Delta \ \XX^j \
U\,,
\label{2.43}\\
&&  \Delta = j  \ , \ \ \ \ \ \ \ \ c_j=\frac{\sqrt{\lambda}}{8 \pi N} \sqrt{j} (j+1) \ ,
\label{2.44}
\eea
where $\XX^j$ is the same as in \rf{2.4} while the 2-derivative  $U$
part is more complicated~\cite{bt2}
\bea
&&
U=U_1+U_2+U_2\,, \ \ \ \ \ \ \ \ \
U_1=\frac{1}{z^2}\big[(\partial_{\alpha}x_m)^2 -(\partial_{\alpha}z)^2\big]-{\cal L}_{S^5}\,,
\la{2.451}\\
&&
U_2=\frac{8}{(z^2 +|x-a|^2)^2}
\Big[|x-a|^2 (\partial_{\alpha}z)^2 - [(x_m -a_m) \partial_{\alpha} x_m]^2\Big]\,,
\nonumber\\
&&
U_3 =\frac{8 (|x-a|^2 -z^2)}{z (z^2 +|x -a|^2)^2} \ (x_n-a_n)\partial_{\alpha}x_n
\ \partial_{\alpha}z\,.
\label{2.45}
\eea
For simplicity, we will consider the case of the regular 4-cusp  Wilson loop;
the corresponding solution~\eqref{1.4} does not deepend on $S^5$  coordinates so $\XX^j=1$.

% Due to a complicated form of~\eqref{2.45}
%calculations for an arbitrary insertion $a_m$ are difficult. Fortunately, all we need to find
To find  the function $F(\zeta)$ in \rf{1.15}   it is sufficient, as in section 3.1.3, to
 choose the special case of  $a=(a_0, 0,0,0) .$
  Remarkably, in this case  $U$  can be put into the following
 simple  form (cf. \rf{2.34})
\bea
    &&U_{{a=(a_0, 0,0,0)}} =\frac{2}{\cosh^2 u \ \cosh^2 v} \frac{1+\varrho^2 -
(\sinh u\ \sinh v + \varrho\ \cosh u\ \cosh v)^2}{(1+\varrho\ \tanh u\ \tanh v)^2}\,,
\label{2.46}
\\
&&\varrho\equiv \frac{2 a_0}{1-a_0^2}\,.
\label{2.47}
\eea
Substituting it into~\rf{2.3},\eqref{2.43} gives
\bea
&&{\cal C}_j(W_4^{(reg)}, a_0)=\frac{2 c_j}{(1-a_0^2)^j} \ \int_{-\infty}^{\infty}d u dv\
\Big[\frac{(\cosh  u\  \cosh v )^{-1}}{1+\varrho\ \tanh u\  \tanh v}\Big]^{j+2}
\nonumber\\
&&\qquad\qquad \qquad\qquad\qquad \qquad \times \Big[1+\varrho^2 -\big(\sinh u \ \sinh v + \varrho\ \cosh u\
\cosh v\big)^2\Big]
\label{2.48}
\eea
For an arbitrary $j$ this integral is rather complicated  but can be easily done
for specific values of $j$. For instance, for $j=2$ we obtain:
\be
{\cal C}_2(W_4^{(reg)}, a_0)=\frac{4 c_2}{3(1-a_0^2)^2 \varrho}\ \log\frac{\varrho+1}{\varrho-1}\,.
\label{2.49}
\ee
To compute $F(\zeta)$ we have to multiply~\eqref{2.49} by the factor (see \rf{1.15})  \
\be \big( |x^{(1)} -  x^{(3)}| \  |x^{(2)} -  x^{(4)}|\big)^{-1}  \
\prod_{i=1}^4 |a- x^{(i)} |=\ { 1 \ov 8}  [(a_0-1)^2 +2] [(a_0+1)^2 +2] \,,
\label{2.50}
\ee
where the positions of the cusps $x^{(i)}$ are given by~\eqref{2.16} and
 we used  that
$\Delta=j=2$.
Expressing  $\varrho$ in terms of $\zeta$ in \rf{1.13}, \rf{1.14}.
 gives  the  same relation as in~\eqref{2.35}.
As a result, we  find  (cf. \rf{2.17})\foot{Let us note that the same  (up to a constant factor)
 expression
 is found by formally setting $\D=2$ in \rf{2.31},\rf{2.33}.}
\be
j=2: \ \ \ \ \ \ \ \ \ \ \ \ \ \ \ \ \ \ \ \ \ \ \ \
F(\zeta)=\frac{ c_2}{3}\ \frac{\sqrt{\zeta} }{\zeta-1}\ \log \z  \,.
\label{2.51}
\ee
We  can then restore the dependence of the  correlator on  all 4 coordinates of $a_m$ getting
the following analog of \rf{2.8}
%The result for the correlation function~\eqref{1.1} can now be found from the general
%expression~\eqref{1.15} for $\Delta=2$. Explicitly, we obtain
\be
{\cal C}_2(W_4^{(reg)}, {a})=-\frac{c_2\ {\log \z}}{3 (\R a_0 -2 a_1 a_2)}\,,
\label{2.52}
\ee
where $\R$ is given by~\eqref{1.14} and $\z$ is defined in~\eqref{1.13}.

Taking the position  of the operator $a$ to infinity
%along, say, $a_3$ direction
we get the corresponding  OPE coefficient
%corresponding
%to the operator conjugate to~\eqref{2.43} in the expansion of the Wilson loop
(we restore the factor of the scale $\el$  of the  loop)
%For this we take the operator far away from the loop along the space-like
%$a_3$ direction and find the following behavior
%
\be
{\cal C}_2(W_4^{(reg)}, {a})_{|a|\to \infty} =  \frac{8 c_2\  \el^2}{3 \ |a|^4}\,.
\label{2.53}
\ee
Here the  power of $|a|$   reflects the value of the dimension $\D=2$ (cf. \rf{2.18}).
We may also  study  the limit when $a_m$  is approaching the position of
one of the cusps.  Choosing  $a_m$ as in~\eqref{1.17} and expanding
for small $\ep$ gives (cf. \rf{2.28.2})
\be
{\cal C}_2(W_4^{(reg)}, {a}) \to
\frac{1}{3 \ep^2}\
\frac{\log \frac{\a_1^2 +\a_2^2 +\a_3^2-\a_0^2}{2(\a_0-\a_1)(\a_0-\a_2)}}
{ -3 \a_0^2+ \a_3^2 +(\a_1-\a_2)^2 +2 \a_0 (\a_1+\a_2) }\,.
\label{2.54}
\ee
%

%%%%%%%%%%%%%%%%%%%%%%%%%%%%%%%%%%%%%%%%%%%%%%%%%%%%%%%%%%%%%%%%%%%%%%%%%%%%%%%%%%%%%%%%%%%%%%%%%%%%%%%%%%%%%%%%%%%%%%%%%%%%

\section{Comment  on relation    to  3-point correlator  with
 two  infinite spin  twist-2 operators  }

%%%%%%%%%%%%%%%%%%%%%%%%%%%%%%%%%%%%%%%%%%%%%%%%%%%%%%%%%%%%%%%%%%%%%%%%%%%%%%%%%%%%%%%%%%%%%%%%%%%%%%%%%%%%%%%%%%%%%%%%%%%%%%

As is well known, the   anomalous  dimension of the large  spin $S$ twist-2 operator
(the coefficient of the  $\ln S$ term in it)  is closely related to the  UV   anomaly
in the expectation value of a Wilson  loop with a null cusp \ci{kor,Martin,AM1,me,AMtwo,refs}.
At strong coupling,  this relation can be understood \ci{KRTT}  by relating the
corresponding string world surfaces by a world-sheet  euclidean  continuation and an $SO(2,4)$
conformal transformation. This may  be effectively interpreted as a  relation between
the 2-point function of twist-2 operators $\langle  \OO^\dagger_S (x)  \OO_S (0)
\rangle $ with   $S\to \infty$
 and the singular part of the  expectation of the  cusped Wilson loop
$\langle W_4 \rangle $.

Below we shall discuss if  such a  relation may apply
also if one includes in the respective
correlators   an extra  ``light''  operator $\OO_\D$ \  ($ \D \ll \sql$),
\be
 \CC (W_4,a) =
\frac{\langle W_4\ {\cal O}_\D ({a})\rangle}{\langle W_4 \rangle}\,   \ \ {\rm vs.}   \ \ \
 K (\xx^{(1)}, \xx^{(2)}, a)=   \frac{\langle {\cal O}_S^{\dagger} (\xx^{(1)} )\ {\cal O}_S(\xx^{(2)} )\ {\cal O}_\D({a})\rangle}
{\langle {\cal O}_S^{\dagger}(\xx^{(1)} )\ {\cal O}_S(\xx^{(2)} ) \rangle}\,.
\label{3.1}
\ee
This relation may be expected  only in the {\it infinite} spin limit  and
one is   to assume a   certain correspondence  between  the
locations of the 4 cusps  $x^{(i)}$ in $W_4$  and the  positions $\xx^{(1)},\xx^{(2)}$
of the twist-two operators.

The reason why this relation may be  expected
 at strong coupling is the following.
 Computed at strong coupling with a ``light'' dilaton operator
 both correlators in \rf{3.1}
are given by  the   vertex operator  corresponding  to $\OO_\D$
evaluated  on the semiclassical world  surfaces  associated, respectively,
 with  $\langle W_4 \rangle$
and with
 $\langle {\cal O}_S^{\dagger}(\xx^{(1)} )\ {\cal O}_S(\xx^{(2)} ) \rangle$.
  These two surfaces  are   closely related \ci{KRTT,bt1}.
As we shall see below, the integrands  for the corresponding semiclassical
correlators will be    the  same but the integration regions, however,
will  differ.  For a special choice of
the location of the ``light'' operator  the results  for the two integrals
will be the same up to a factor of 2.

Indeed, the  solution~\eqref{1.4} is essentially the same as
the semiclassical trajectory
supported by the two twist-2 operators in the large spin limit
$S \gg \sqrt{\lambda}$
%~\cite{bt1}
%
\bea
&& z=\frac{1}{\cosh (\kappa \tau_e)\ \cosh (\mu \s)}\,, \quad
x_0= \tanh (\kappa \tau_e)\ \tanh (\mu \s)\,, \nonumber\\
&&
x_1= \tanh (\kappa \tau_e) \,, \quad x_2= \tanh (\mu \s) \,, \quad x_3=0\,,
\label{3.3}\\
&&\k= \mu \gg 1  \ , \ \ \ \ \ \ \
 \kappa =\frac{\Delta_S-S}{\sqrt{\lambda}}\,, \qquad \mu =\frac{1}{\pi}\log S\,.
\label{3.3.0}
\eea
The  relation
$\k=\m$ follows from the Virasoro
condition which is also implied by the marginality of the twist 2 operator.
This  background is  equivalent to   the  one
 found  by a euclidean rotation ($\tau \to i \tau_e$)  of the  large spin limit
of the folded spinning string in $AdS_5$.\foot{
To find the
 solution with the  singularities
prescribed by the two twist-2 operators one has to
act on~\eqref{3.3} with a two-dimensional conformal
map which sends the cylinder to the plane
with two marked points, see ~\cite{bt1,b} for details.}
%
%is exactly the same as the semiclassical trajectory supported by two twist-2 operators
%in the limit of large spin $S \gg \sqrt{\lambda}$~\cite{BT1}. Nevertheless, there is an important
%difference. Twist-2 operators are dual to a closed string state spinning in $AdS_3$. However, the
%solution~\eqref{3.3} for $u, v \in (-\infty, \infty)$ has open worldsheet. To form a closed
%worldsheet with topology of a cylinder we have to treat one of the variables, say $v$, as periodic.
%The obtained solution coincides with the Euclidean version of the folded string solution
%of~\cite{GKP} written in Poincare coordinates.
Here $\tau_e \in (-\infty, \infty)$. In the original  closed string solution
  $\sigma \in [0,2\pi]$;
in fact, in  \rf{3.3}  we have    $\sigma \in [0, {\pi\ov 2}]$
and  3 other segments are assumed to be added similarly.
In the  infinite spin   limit $\mu \to \infty $ so
we may formally set $v= \mu \s \in (0, \infty)$.
Introducing also $u= \kappa \tau_e\in (-\infty, \infty)$ we conclude
that \rf{3.3} becomes  equivalent to \eqref{1.4}, up
%AT
to the ``halved''  range of  $v$.

More precisely, the operators considered in ~\cite{bt1} were  defined on a  euclidean
4-space and the corresponding  surface  had imaginary $x_2$; the real
surface  equivalent  to \rf{3.3} is obtained by the Minkowski  continuation in
the target space $x_2 \to i x_0, \ x_0 \to x_2$.
Below we shall interchange the notation
 $x_1 \leftrightarrow x_2$  compared to ~\cite{bt1}.
 Since in ~\cite{bt1} the
operators were assumed to be located at $R^4$ points $\xx^{(1,2)} =(\pm 1, 0, 0, 0)$
 taking this continuation into the  account
    we conclude that
the solution~\eqref{3.3} is the semiclassical
trajectory  saturating  the two-point correlator of twist-2 operators
located at the following special points in the
Minkowski 4-space\foot{Note that
if  $\mu \s$ in~\eqref{3.3} were not extending  to infinity, the boundary behavior of~\eqref{3.3}
would be  different
from a rectangular Wilson loop. The boundary would be  reached only for $u \to \pm \infty$
and on the boundary we would  have a piece of the null line $x_1=x_0$ located
at $x_2=1$ and a piece of the null line $x_1=-x_0$ located at $x_2=-1$. These null lines
would  no longer be connected on the boundary.}
\be
\xx^{(1)} =(0, 1, 0, 0)  \ ,\ \ \ \ \ \ \ \ \ \ \xx^{(2)} =(0, -1, 0,0)  \ .
\label{3.3.1}
\ee
%
%\iffalse
%In order to compute
%$\frac{\langle {\cal O}_S^{\dagger} {\cal O}_S {\cal O}_{dil}({a})\rangle}
%{\langle {\cal O}_S^{\dagger} {\cal O}_S \rangle}$
%in the limit when the dimension of the operator is much less than $S$ we use the same procedure
%as in the previous section. That is we evaluate the vertex operator~\eqref{2.4} on the solution~\eqref{3.3}.
%We refer to~\cite{RT, BT2}
%for details of calculations involving twist-2 operators.
%Here we simply quote that in the limit of strictly infinite spin
%the upper limit of the integral over $\s$ is extended to infinity.
%The coefficient of the three-point
%function $\langle {\cal O}_S^{\dagger} {\cal O}_S {\cal O}_{dil}({a})\rangle$ is
%finite in this limit~\cite{RT}.
Assuming for definiteness that $\OO_\D$ is the dilaton operator
we conclude that   the leading strong-coupling term
in the  correlator of two infinite  spin   twist-2 operators and  the dilaton
$K_{dil} =\frac{\langle {\cal O}_S^{\dagger} (\xx^{(1)}) \ {\cal O}_S (\xx^{(2)})\ {\cal O}_{dil}({a})\rangle}
{\langle {\cal O}_S^{\dagger}(\xx^{(1)})\  {\cal O}_S(\xx^{(2)}) \rangle}$
%in the limit $S \to \infty$ has to be
%given (up to a coefficient\footnote{The precise numeric coefficient can be
%restored using details of
%calculations in~\cite{RT, BT2}.
%It is not important for our purposes.})
is then given by  the same integral as in~\eqref{2.7} with the  only difference  being
that instead of  the range of integration
$v \in (-\infty, \infty)$ that  we had   in the Wilson loop case
now
in the closed string case  we have $v \in [0, \infty)$  with the whole
integral  multiplied by 4.
Equivalently, the spatial
integral for the folded string solution is done over $\s \in [0, {\pi \ov 2})$
with the result multiplied by 4  \ci{rt}.
The topology
of the world sheet parametrized by $(u,v)$  should be a disc (or a  plane with infinity removed)
in the Wilson loop case  and the cylinder (or a half-plane with infinity removed)
in the folded closed string case.

Note that if we set $a_0=a_2=0$ in the integrand in~\eqref{2.7} it becomes an
even function of $v$ and thus the  integral over $v\in [0, \infty)$
is just half the  integral over $v \in (-\infty,\infty)$.
%This implies that~\eqref{2.8} has to be consistent with \rf{3.44}  in this
%special case.
Thus for the special values of $a$ the two correlators  in \rf{3.1} are directly related.

For  general $a$ we get for $K_{dil}$  an expression  that is
different from  \rf{2.8}
% here we get
%The result for the integral in this case is as follows
%
\bea
&& K_{dil}= \frac{c_{dil} }{{3 (\R^2- 4 a_1^2)(\R a_0- 2 a_1 a_2)^3}} \Big[
-4\big[\R^2-4a_1(a_0+a_1) +2\R a_2\big] (\R a_0 -2 a_1 a_2) \no \\
&& + (\R^2- 4 a_1^2) (\R^2+4a_0^2 -4 a_1^2 -4 a_2^2)
\log \frac{ (\ha \R+ a_1) (\ha \R+ a_0 - a_1 - a_2)}{(\ha \R- a_1)(\ha \R- a_0 + a_1 - a_2)}
\Big]\,.
\label{3.3.2}
\eea
This expression  containing the  logarithm of a coordinate ratio
appears to be in conflict with the expected  structure of the 3-point function
of conformal primary operators ($x^{(ij)}=x^{(i)} - x^{(j)}$)
\be
\langle {\cal O}_{\Delta_1} ({{\rm x}}^{(1)})
{\cal O}_{\Delta_2} ({{\rm x}}^{(2)})
{\cal O}_{\Delta_3} ({{\rm x}}^{(3)})
\rangle=
\frac{C_{123}}{|{{\rm x}}^{(12)}|^{\Delta_1+\Delta_2-\Delta_3}
|{{\rm x}}^{(23)}|^{\Delta_2+\Delta_3-\Delta_1}
|{{\rm x}}^{(31)}|^{\Delta_3+\Delta_1-\Delta_2}
}\,.
\label{3.4}
\ee
In  the present case with $\Delta_S \gg \Delta_{dil} =4$ and $a={{\rm x}}^{(3)}$
i\rf{3.4} can be explicitly  written as
\be
\frac{\langle {\cal O}_S^{\dagger} (\xx^{(1)})\  {\cal O}_S
(\xx^{(2)})\ {\cal O}_{\D}({a})\rangle}
{\langle {\cal O}_S^{\dagger}(\xx^{(1)})\  {\cal O}_S(\xx^{(2)}) \rangle}
\  \approx  \
\frac{C}{
|{{\rm x}}^{(1)} - a|^{\D} |{{\rm x}}^{(2)}- a|^{\D}
}\,,
\label{3.44}
\ee
where  $C$ is  a  constant which is  finite  in the $S \to \infty$ limit \ci{rt}.

An  explanation of this apparent puzzle is as follows.
The standard argument leading to \rf{3.4} assumes
that % which allow one to fix three-point function in the form~\eqref{3.4}
%rely on the fact that the operators
the 3  points $\xx^{(i)}$ can be spatially separated (as is always the case in $R^4$ but not in $R^{1,3}$  we consider here).
%For this, correlation functions are
%usually considered in Euclidean field theory.
% However, twist-2 operators are intrinsically non-Euclidean
%and the positions of the operators has to be chosen with some care.
If we
choose the dilaton operator insertion point $a$  to be   away from the
$(x_0,x_2)$-plane  ($(x_1,x_2)$ plane of rotation of the original spinning  string)
we can spatially
separate all the operators.  For example, we may set  $a_0=a_2= 0$.
In this case  the integrand in~\eqref{2.7}  becomes an
even function of $v$ and the integral over $v\in [0, \infty)$
is just half the  integral over $v \in (-\infty,\infty)$, i.e.
%This implies that
 ~\eqref{2.8} should be  proportional to  \rf{3.44}.
%special case.
 Indeed,  taking the limit   $a_0=a_2= 0$
in ~\eqref{2.8}, \rf{3.44}   we find that  the logarithmic term   goes away  and
we obtain
\be
{\cal C}_{dil}( W_4^{(reg)}, a)_{_{a_0=a_2=0}}= \ha  (K_{dil}){_{{a_0=a_2=0}}}
 =
 \frac{32 c_{dil}}{9}\  \frac{1}{[(a_1-1)^2+a_3^2]^2\ [(a_1+1)^2+a_3^2]^2} \,.
\label{3.5}
\ee
This is  consistent with \rf{3.44} if we recall the expression \rf{3.3.1}
for the locations of the twist-2 operators.

Similar analysis can be repeated for the correlation function with the chiral primary operator
with $\D=j=2$.
Taking the limit
 $a_0=a_2=0$ in  ~\eqref{2.52}  leads to a similar   expression
\be
{\cal C}_{2}( W_4^{(reg)}, a)_{_{a_0=a_2=0}}= \frac{8 c_2}{3}\frac{1}{[(a_1-1)^2
+a_3^2]\ [(a_1+1)^2 +a_3^2]}\,,
\label{3.6}
\ee
again consistent with \rf{3.4},\rf{3.44}.
%
%This is
%precisely the correct form of the ratio
%$\frac{\langle {\cal O}_S^{\dagger} {\cal O}_S {\cal O}_{2}({a})\rangle}
%{\langle {\cal O}_S^{\dagger} {\cal O}_S \rangle}$.

Finally, let us consider the generalized cusp
 solution with $S^5$  momentum~\eqref{2.36}.
The  generalization of the solution \rf{3.3} to the case  when large spin operators carry
also a large  angular momentum $J$  in $S^5$ is   given by
\bea
 &&z=\frac{1}{\cosh (\kappa \tau_e)\ \cosh (\mu \s)}\,, \qquad
x_0= \tanh (\kappa \tau_e)\ \tanh (\mu \s)\,, \no\\
&&x_1= \tanh (\kappa \tau_e) \,, \qquad x_2= \tanh (\mu \s) \,, \qquad x_3=0\,,
\la{3.6.1}\\
&&
\varphi=-i \nu \tau_e \,, \qquad \nu =\frac{J}{\sqrt{\lambda}}\ ,\
\ \ \ \  \ \
\kappa^2 =\mu^2 +\nu^2\,.
\la{3.6.2}
\eea
Taking  the scaling limit \ci{ft}
when $\k,\mu,\nu \to \infty$   with fixed
\be
k\equiv  {\kappa\ov \mu}\ , \ \ \ \ \ \ \ \ \ell \equiv  {\nu \over \mu}= {\pi J \ov \sql \ln S} \
, \ \ \ \ \ \ \ \  k^2 =1 + \ell^2 \ ,
%\ \ \ \ \ \ \mu \to \infty \ ,
\la{3.8}
\ee    and setting
$v = \mu \s\in [0, \infty), \   u= \mu \tau_e\in (-\infty, \infty)$
% (which preserves the canonical form
%of the flat metric in conformal gauge)
 we obtain  the background \rf{2.36}, again
up to a different range of $v$.
Taking the   limit   $a_0=a_2=0$ in      ~\eqref{2.41} with $\ell\not=0$
 gives
\be {\cal C}_{dil}( W_{4,\ell }^{(reg)}, {a})_{_{a_0=a_2=0}}=
\frac{32 c_{dil}}{9 \sqrt{1+\ell^2}}\  \frac{1}{[(a_1-1)^2+a_3^2]^2\ [(a_1+1)^2+a_3^2]^2} \,.
\label{3.7}
\ee
Comparing with the result ~\cite{rt} of the semiclassical computation of the
corresponding correlator
$\frac{\langle {\cal O}_{S, J}^{\dagger} {\cal O}_{S, J} {\cal O}_{dil}({a})\rangle}
{\langle {\cal O}_{S, J}^{\dagger} {\cal O}_{S, J} \rangle}$
we find again the agreement: the 3-point function coefficient
in eq. (4.18) of \ci{rt} also scales as  ${1 \ov \sqrt{1+\ell^2}}$ in the
limit \rf{3.8}.
%provided $\ell =\nu/\mu$ where $\mu$ and $\nu$ are defined in~\eqref{3.3.0}
%and~\eqref{3.3.01}. In orther words, eq.~\eqref{3.7} represents the three-point function
%in the limit
%
%\be
%\frac{J}{\sqrt{\lambda}} \to \infty \,, \quad S \to \infty \,, \quad
%\ell =\frac{\pi J}{\sqrt{\lambda} \log S} \quad{\rm fixed}\,.
%\label{3.8}
%\ee
%

The above discussion  establishes a certain
 relation between the two correlators in \rf{3.1}
  at strong coupling. An open  question is whether
a similar relation may hold also  at weak  coupling.
Another natural question is whether this relation  may generalize
to  correlators involving $W_n$ with  $n>4$ and  the corresponding  number of
large spin twist-2 operators.

%%%%%%%%%%%%%%%%%%%%%%%%%%%%%%%%%%%%%%%%%%%%%%%%%%%%%%%%%%%

%%%%%%%%%%%%%%%%%%%%%%%%%%%%%%%%%%%%%%%%%%%%%%%%%%%%%%%%%%%%%%%%%%%%%%%%%%%%%%%%%%%%%%%%%%%%%%%%%%%%%%%%%%%%%%%%%%%%%%%%%

\section{On  correlator of Wilson loops with   $n > 4$  cusps and dilaton  at strong coupling }

%%%%%%%%%%%%%%%%%%%%%%%%%%%%%%%%%%%%%%%%%%%%%%%%%%%%%%%%%%%%%%%%%%%%%%%%%%%%%%%%%%%%%%%%%%%%%%%%%%%%%%%%%%%%%%%%%%%%%%%%%

For   number of cusps $n >4$  there are no explicit solutions  like~\eqref{1.4}
or~\eqref{2.19}  known so far.  This makes it  difficult to generalize the analysis
of the previous sections to  $n>4$. Below we will calculate numerically
the large distance ($|a|\to \infty$)  limit of the correlator \rf{1.1},\rf{2.3}  with an  even number of cusps $n$ at strong coupling,
i.e.
the corresponding OPE coefficient.
% corresponding to the dilaton in the expansion
%of the regular Wilson loop with an  even number of cusps.

Minimal surfaces ending on regular polygons with an even number of
cusps were studied in \cite{AM2}, to which we refer the reader for details.
Such surfaces can be embedded in $AdS_3$
and are described, using the Pohlmeyer reduction of the classical $AdS_3$ string equations,
 in terms of the two reduced-model  fields:
a holomorphic function $p(\xi)$ and a field $\alpha(\xi,\bar \xi)$ satisfying the generalized sinh-Gordon equation
\begin{equation}
\label{sinh-gordon}
\partial \bar \partial \alpha +p \bar p\  e^{-2\alpha}-e^{2\alpha}=0\,.
\end{equation}
Here $\del = {\del \ov \del \xi}$  and  $\xi$ is a complex coordinate on the world sheet $\xi= u+i v$.
In order to reconstruct  the  corresponding solution of the  original string equations
  one is to solve the following two matrix  auxiliary linear problems
  ($\b= (\xi,{\bar \xi})$) %(denoted as left and right)
    \begin{eqnarray}\la{lin}
(\del_\b +B^{L}_\b)\psi_{L} =0\,, \ \ \ \ \ \ \ \ \  (\del_\b +B^{R}_{\b})\psi_{R} =0\,, \
\end{eqnarray}
%
%The flat connections are given by
%
where the  components of the flat connections are
\begin{eqnarray}
B^L_{\xi}= \left( \begin{array}{cc}
\ha  \partial \alpha & -e^\alpha \\
-e^{-\alpha}p(\xi) & -\ha \partial \alpha \end{array} \right),~~~\ \ \ B^L_{\bar \xi}=\left( \begin{array}{cc}
-\ha  \bar \partial \alpha & -e^{-\alpha}\bar p(\bar \xi) \\
-e^\alpha & \ha  \bar \partial \alpha \end{array} \right),\\
B^R_{\xi}= \left( \begin{array}{cc}
-\ha  \partial \alpha & e^{-\alpha}p(\xi) \\
-e^{\alpha} & \ha  \partial \alpha \end{array} \right),~~~B^R_{\bar \xi}=\left( \begin{array}{cc}
\ha  \bar \partial \alpha & -e^\alpha  \\
e^{-\alpha} \bar p(\bar \xi) & -\ha  \bar \partial \alpha \end{array} \right)\,.
\end{eqnarray}
The flatness of these connections is equivalent to  eq. (\ref{sinh-gordon})
and the  holomorphicity of $p(\xi)$. The solutions describing   regular polygons
correspond to the holomorphic function being a homogeneous polynomial of degree $k-2$, with $n=2k$ and $\alpha$
being a function of the radial coordinate  $|\xi|=\sqrt{\xi \bar \xi} $ only, i.e.
\begin{equation}
p(\xi)=\xi^{k-2}\,,~\ \ \ \ \ \ \ ~~\bar p(\bar \xi)=\bar \xi^{k-2}\,,~~ \ \ \ \ \ ~\alpha=\alpha(|\xi|)\,.
\end{equation}
In this case the sinh-Gordon equation \rf{sinh-gordon} reduces to the  Painleve III  equation.
The boundary conditions are that $\alpha$ is regular everywhere and $\hat \alpha = \alpha-{1 \ov 4}  \log p \bar p$
vanishes at infinity.
The solution to this equation can be written in terms of Painleve transcendentals.

While for computing the area of the world-sheet we only need to know the reduced fields,
in order to compute the correlation function with the
dilaton  according to \rf{2.3} we need an explicit expression for the space-time coordinates.
%While  we do not know how to solve the auxiliary linear problem in general,
The general strategy  of finding  the solution  for the  string space-time coordinates is as follows.
%AT-- Fernando could you clarify this -- psi are not matrices so phasing is not good
One should first find
 the solutions $\psi_{L},\psi_R$.
% are $ 2 \times 2$  matrices
%that are chosen in such a way that they are equal to the identity at $|\xi|=0$
%(different choices are related by conformal symmetry).
Then the solution for the
  string $AdS_3$  embedding  coordinates  $Y_M$ is given by
\begin{equation}
{\rm Y}_{a,\dot a} =( \psi_L^T)_a (\psi_R)_{\dot a}  =  \left( \begin{array}{cc}
Y_{-1}+i Y_0&Y_1-i Y_2 \\
Y_1+i Y_2& Y_{-1}-i Y_0  \end{array}\right)\,.
\end{equation}
The form of the  solution in Poincare coordinates  is determined by
\begin{eqnarray}
z= \frac{1}{Y_{-1}},~~~\ \ x_0=\frac{Y_0}{Y_{-1}},~~~ \ \ x_1=\frac{Y_1}{Y_{-1}},~~~ \ \ x_2=\frac{Y_2}{Y_{-1}} \,.
\la{77}
\end{eqnarray}
Unfortunately, we do not know  how  solve the above linear problems for $\psi_{L,R}$, i.e.
how to compute  the coordinates in \rf{77}  and thus  the integral in \rf{2.3}  for $n > 4$. \footnote{In \cite{amf,Bubble} and subsequent developments (see \cite{Alday:2010kn} for a review) it was understood how to use integrability in order to compute the area of the world-sheet even without knowing its shape. It would be interesting to learn how to apply similar tricks to solve the problem at hand. On the other hand, following the methods of \cite{Gaiotto:2011tf} one could try to construct the shape of the world-sheet and then  
%Ark
 solve the current problem.}

For that reason in  this paper we shall  focus on the
simpler problem of computing the large distance limit $|a| \to \infty$ of the correlation function
or the OPE coefficient.
 % the OPE coefficient, that is
%the correlation function in the limit in which the operator
%is inserted at a very large distance. The operators will be chosen
%to be the dilaton.
In this limit the integral in ~\eqref{2.4} takes a simpler form
(here $\XX^j=1$; cf. \rf{1.16})
\bea
&& \C_{dil} (W_n, a)_{|a|\to \infty}=  \frac{\rC_n }{|{a}|^{2 \Delta}} \ ,
\ \ \ \ \ \ \ \ \ \ \rC_n=   c_{dil}\  I_{n,\Delta} \ , \la{444}  \\
&& I_{n,\Delta}=  8 \int d u d v \  z^\Delta \    e^{2\alpha}  =
8
\int du dv \ \frac{e^{2\alpha}}{Y_{-1}^\Delta}\,,
\label{B0}
\eea
where $\Delta=4+j$\  (here we keep  the angular momentum  $j$  of the dilaton
general).
We used the fact that in the Pohlmeyer reduction the string  Lagrangian or $U_{dil}$ in \rf{2.4}
 is given by
$8 e^{2\alpha}$.
The  integral  \rf{B0} (and, in fact, the full integral in \eqref{2.4})
 can be computed exactly in the two cases:  (i) $n=4$ we already discussed above (see
\rf{2.31})    and (ii)
 $n\to \infty$  when the cusped polygon  becomes a  circular Wilson loop.
%Using~\eqref{2.4} and the solution~\eqref{1.4}, we obtain for $n=4$
%
In the $n=4$  case we get using ~\eqref{2.4},\eqref{1.4}    (cf. \rf{2.1},\rf{2.31})
\begin{equation}
I_{4,\Delta}  = {2 \pi } \  \Big(\frac{\Gamma[{\Delta\ov 2}]}{\Gamma[{\Delta+1 \ov 2}]} \Big)^2\,.
%~~~~~I_{\infty,\Delta}= \frac{\pi}{2(\Delta-1)}
\label{n=4}
\end{equation}
The  world-surface ending on a circular loop  is given (in conformal gauge) by
%For the case of the circle we can also use~\eqref{2.4} and evaluate it (in the limit
%of large $|{a}|$) on the circular Wilson loop solution
~\cite{cor}
\bea
&&z=\tanh u\,, \qquad x_1 =\frac{\cos v}{\cosh u}\,, \qquad
x_2 =\frac{\sin v}{\cosh u}\,, \qquad x_0=x_3=0\,; \nonumber\\
&&u\in [0, \infty)\,, \quad v\in [0, 2 \pi]\,.
\label{cir}
\eea
This leads to
\be
I_{\infty,\Delta}= \frac{4 \pi }{\Delta-1} \ \,.
\label{cir2}
\ee
For the case of  general even $n$   and $j=0$, i.e.
$\Delta=4$  we can compute the  integral  \rf{B0}  numerically.
The numerics is  very accurate since the
 factor $z^4$ makes the integrand decay very fast.
Let us  simply present the results for  $I_n \equiv I_{n, \D=4}$ in a few cases
%(ignoring the obvious factors $c_{dil}/|{a}|^8$)
%
\begin{eqnarray}
I_4={ 32\ov 9}& \approx & 3.55556\no \\
%0.444444\\
I_6& \approx &3.90901%\la{six}
\no  \\
%0.488626 \\
I_8& \approx & 4.04496\no \\
%0.50562\\
I_{10}& \approx &4.10648\no \\
%0.51331 \\
I_{\infty}=\frac{4 \pi}{3} & \approx & 4.18879\,,
%0.52359877
\la{six}\end{eqnarray}
where we have also included the two special cases
discussed above  when $I_n$ is   known exactly.
Unfortunately, we could not find  analytic expressions in agreement with these
numbers.\footnote{$I_6$ is well approximated by $4 \sqrt{\frac{3}{\pi}}$.}

%%%%%%%%%%%%%%%%%%%%%%%%%%%%%%%%%%%%%%%%%%%%%%%%%%%%%%%%%%%%%

\section{Correlation function of cusped Wilson loop with\\ dilaton operator  at weak coupling}

Let us now consider the computation of the correlator \rf{1.1}
\be
{\cal C}(W_n, {a})=\frac{\langle W_n {\cal O} ({a})\rangle}{\langle W_n\rangle}\,,
\label{6.1}
\ee
 in the weakly coupled planar $SU(N)$
${\cal N}=4$  supersymmetric  gauge  theory. Here the expectation values are computed
using gauge theory path integral  and\foot{The additional coupling to the scalars  in  the locally-supersymmetric  Wilson loop
 \cite{Juan} drops out because the null polygon   contour
consists of null lines.}
% At week coupling the Wilson loop is given by %~\cite{Juan,RY}
%
\be
W_n =\frac{1}{N}{\rm tr}\ {\cal P}\ e^{i g \oint_{\gamma}A_{m}dx^{m}}\,,
\label{6.2}
\ee
 Here we rescaled the fields with the coupling constant $g$   (with $\l = g^2 N$)
so that the ${\cal N}=4$ Lagrangian is
\be
{\cal L}_{{\cal N}=4}=-\frac{1}{4}{\rm tr}(F_{mn}^2 + \dots)
\label{6.3}
\ee
with  $g$ appearing  only in the vertices. We use the  conventions
\be
A_{m}=A_{m}^r T^r\,, \qquad {\rm tr}(T^r T^s)=\delta^{r s}\,, \qquad r, s =1,\dots, N^2-1\,.
\label{6.3.1}
\ee
The path $\gamma$ in~\eqref{6.2} is the union of $n$ null segments of the form
\be
\gamma^{(i)}_{m} (t)=x^{(i)}_{m} + t (x^{(i+1)}_{m}-x^{(i)}_{m})\,,\ \  \qquad t \in [0, 1]\,,
\label{6.3.2}
\ee
where $x^{(i)}_{m}$  ($i=1,...,n$) denote the locations of the  cusps.
%AT
The dilaton operator (which  is a supersymmetry descendant of $\tr Z^2$)
is essentially
the ${\cal N}=4$ gauge theory  Lagrangian up to a total derivative (see, e.g., ~\cite{KTvR})\foot{
%A27
Up to the  scalar and the fermion equation of motion terms ${\cal O}_{dil}$ is thus given by the 
YM Lagrangian plus the Yukawa and the quartic scalar interaction terms.}
\be
{\cal O}_{dil}=\ \hat{c}_{dil}\ {\rm tr}(F_{mn}^2 +\Phi^I \partial^2 \Phi^I +\bar \psi \gamma \cdot
\partial \psi +\dots)\,,
\label{6.4}
\ee
where $\Phi^I$ are the scalars and $\psi$  are  the  fermions  and we did not write
explicitly  the terms
of order $g$ and $g^2$.
The normalization coefficient $\hat{c}_{dil}$ is given by~\cite{Liu}
\be
\hat{c}_{dil}=\frac{\pi^2}{4 \sqrt{3} N}\,.
\label{6.5}
\ee
The leading order contribution to \rf{6.1}
  (to which we will refer  as the ``tree level'' one)  is proportional to  $g^2$
as one can easily see from~\eqref{6.1}, \eqref{6.2}. To compute~\eqref{6.1} to this order we
 have to expand $W_n$ to order $g^2$. Hence, we can set $g=0$ in the Lagrangian~\eqref{6.3}
and in the dilaton operator~\eqref{6.4}. Therefore, for the  purpose of computing the leading order term
in \rf{6.1}  we can take
\be
{\cal O}_{dil} \ \to \ \hat{c}_{dil}\ {\rm tr} F_{mn}^2=
2 \hat{c}_{dil}\  (\partial_{m} A_{n}^r \partial^{m} A^{n r} -
\partial_{m} A_{n}^r \partial^{n} A^{m r})\,.
\label{6.6}
\ee
The gluon propagator in the above  conventions is
\begin{equation}
\langle A_{m}^r({x}) A_n^s(0) \rangle =
-\frac{1}{4\pi^2 {|x|}^2}\ \eta_{mn} \delta^{r s}\,.
\label{6.7}
\end{equation}
We will  see that just like at strong coupling, the weak coupling correlator ~\eqref{6.1} is
finite, i.e. we do not  need
to introduce a UV regularization in~\eqref{6.7}. Also note that to compute~\eqref{6.1}
to order $g^2$ we can replace $\langle W_n \rangle$ in the denominator with unity. Therefore,
we obtain
\bea
&&{\cal C}^{(g^2)}_{dil}(W_n, {a})=
\langle W_n {\cal O}_{dil}({a})\rangle_{tree}\nonumber \\
&&
=-\frac{2 \hat{c}_{dil}\ g^2}{N}
\langle {\cal P} \oint A_k^s({x}) dx^k \oint A_{l}^s({x}')dx'^l
(\partial_p A_q^r \partial^p A^{q r}- \partial_p A_q^r \partial^q A^{p r})({a}) \rangle \ .
\label{6.8}
\eea
The path ordering  symbol ${\cal P}$ means that ${x}'$ in
the second integral is placed between the origin (an arbitrary point along the loop, for instance one of the cusps) and ${x}$.
Now using that
\begin{equation}
\langle A_{k}^r({x}) \partial_p A_q^s({a})\rangle
= -\frac{1}{4\pi^2}{\partial \ov \partial {a^p}} \frac{\eta_{k q} \delta^{r s}}{|{a}-{x}|^2} =
-\frac{1}{2\pi^2} \frac{(a-x)_p \eta_{kq} \delta^{r s}}{|{a}-{x}|^4} \ ,
\label{6.9}
\end{equation}
and
performing the Wick contractions we obtain  ($\l=g^2 N$)
\bea
&&{\cal C}^{(g^2)}_{dil}(W_n, {a})=
-\frac{\hat{c}_{dil}\ \l}{\pi^4} \  {\cal P} \Big(
\oint \oint \Big[
\frac{({a} -{x})\cdot({x}-{x}')}{|{a}-{x}|^4|{a}-{x}'|^4}
d {x} \cdot d {x}' \no
\\ && \ \ \ \ \ \ \ \ \ \ \ \ \ \ \ \ \ \ \ \ \  \ \ \ \ \ \ \ \ \ \ \ \ \ \ \ \ \ \ \  \ - \frac{({a} -{x}) \cdot d{x}'}{|{a}-{x}|^4}
\frac{({a}-{x}') \cdot d{x}} {|{a}-{x}'|^4} \Big]\Big) \,.
\label{6.10}
\eea
So far our discussion have been general and applicable for any number of cusps
 $n$. Let us now specify to  $n=4$.
% and consider the  regular Wilson loop whose cusp locations $x^{(i)}_m$
%are given by~\eqref{2.16}.  In this case it is straightforward to
 Computing  the ${\cal P}$-ordered integrals
in~\eqref{6.10}  we obtain for generic locations of 4 null cusps
\be
{\cal C}_{dil}(W_4^{(reg)}, {a}) =-\frac{\hat{c}_{dil}\ \lambda}{2\pi^4 }\
\frac{   |x^{(1)} - x^{(3)}|^2\   |x^{(2)} - x^{(4)} |^2   }{\prod_{i=1}^4 |a-x^{(i)}|^2}\,.
\label{6.11}
\ee
This agrees with the expected structure \rf{1.15} of the  correlator
(for the dilaton  $\D=4$)  with
the leading weak-coupling term in the
 function $F(\zeta)$ thus  being   simply a constant
\be
F(\zeta)=-\frac{ \hat{c}_{dil}\  \lambda}{2\pi^4}\,.
\label{6.12}
\ee
%
%If we choose the Wilson loop to be irregular four-cusped polygon whose cusp locations
%are given in~\eqref{2.20} we  obtain~\eqref{6.11} with the extra factor
%$1/(1-b^2)^2$ and thus the same  $F$ in \rf{6.12}
% in full analogy with our strong coupling results  (cf.  \rf{bbb}).
%This  verifies the general structure of the correlator \rf{1.15} at  weak coupling.
%\subsection{Tree level result for a generic number of edges}
%Similarly, we can also compute the tree level contribution to the
%correlation function of the dilaton with $n$ cusped polygonal Wilson loop
%for any $n$.
%AT
Note that the structure of \rf{6.11} is exactly the same as the one appearing
in the  1-loop correction to the 4-cusp  Wilson loop  $ \lg W_4 \rg$  (given by a scalar box diagram).
Indeed, integrating \rf{6.11} over $a$ we get the integrated dilaton operator or   gauge theory action
insertion into the Wilson loop,  which is  proportional to derivative of $ \lg W_4 \rg$  over gauge coupling
\ci{Drum,kor}.
%AT18
 This observation may allow one  to extract higher order corrections  to \rf{6.11}
by comparing to  integrands of  higher-order corrections  to $ \lg W_4 \rg$.

When computing the analogs of the  integrals in~\eqref{6.10} for $n >4$
we have two different types  of contributions.
The first one is when the two line integrals are taken  along the same segment.
Let us  call this contribution $T_{ii}$ where the $i$-th segment is parametrized
by~\eqref{6.3.2}.
After some computation we obtain (up to the obvious factor $-\frac{\hat{c}_{dil}\ \lambda}{\pi^4}$)
\begin{equation}
T_{ii}({a})=
-\frac{1}{2} \frac{[({a}-{x}^{(i)})\cdot({x}^{(i+1)}-{x}^{(i)})]^2}
{[({a}-{x}^{(i)})\cdot({a}-{x}^{(i)}))^2(({a}-{x}^{(i)})
\cdot(2 {x}^{(i+1)}-{a}-{x}^{(i)})]^2}\,.
\label{6.13}
\end{equation}
The other type  of contribution appears  when the two contractions are made in different segments.
In this case we obtain
\bea
&&
T_{ij}= \nonumber\\
&&
\frac{({a}-{x}^{(i)})\cdot({a}-{x}^{(j)}){\ }
({x}^{(i+1)}-{x}^{(i)})\cdot({x}^{(j+1)}-{x}^{(j)})}
{|{a}-{x}^{(i)}|^2{\ }
|{a}-{x}^{(j)}|^2 {\ }
({a}-{x}^{(i)}) \cdot ({a}+{x}^{(i)}-2 {x}^{(i+1)}) {\ }
({a}-{x}^{(j)}) \cdot( {a}+{x}^{(j)}-2{x}^{(j+1)})}  \nonumber\\
&&
-\frac{({a}-{x}^{(i)})\cdot({x}^{(j+1)}-{x}^{(j)}){\ }
({a}-{x}^{(j)})\cdot({x}^{(i+1)}-{x}^{(i)})
}
{|{a}-{x}^{(i)}|^2 {\ }
 |{a}-{x}^{(j)}|^2 {\ }
({a}-{x}^{(i)}) \cdot ({a}+{x}^{(i)}-2 {x}^{(i+1)}) {\ }
({a}-{x}^{(j)}) \cdot( {a}+{x}^{(j)}-2{x}^{(j+1)})} \,. \nonumber\\
\label{6.14}
\eea
These expressions  are completely general. Hence, the full answer  (which  is rather  lengthy)  will be the sum of
such  contributions.

Let us specify~\eqref{6.13}, \eqref{6.14} to the case of regular polygons with
even $n$ sides  with  the cusps  located at\foot{Note that for $\to \infty$
this null polygon becomes a unit circle in the (12) plane.}
\begin{equation}
x^{(i)} =
\Big((-1)^i \frac{\sqrt{1-\cos{ 2\pi\ov n}}}{1+\cos {2 \pi\ov n}}, {\ }
\frac{\cos({ \pi\ov n}(2i+1))}{\cos { \pi\ov n}},{\ }
\frac{\sin({ \pi\ov n}(2i+1))}{\cos{ \pi\ov n}},{\ } 0\Big)\,. \label{6.15}
\end{equation}
The problem is purely combinatorial, but  there does not seem to be a simple universal
 formula for generic $n$. It is relatively easy, however, to
 compute the OPE coefficient by placing the operator very far from the loop:
taking $|{a}|$ large we obtain (cf. \rf{1.3},\rf{1.16},\rf{444})
\begin{equation}
{\cal C}_{dil}(W_n^{(reg)}, {a}) _{|a|\to \infty} =
\frac{\rC_n}{|{a}|^8}\,, \ \ \ \ \ \ \ \ \   \rC_n=
-\frac{ 2 \hat{c}_{dil} \lambda}{\pi^4}\ n^2 \tan^2{ \pi\ov n}  \ .
\label{6.16}
\end{equation}
%
%we can check, for instance, than in the large $n$
%limit we obtain $n^2 \tan^2(\pi/n) \rightarrow \pi^2$, which agrees with the result for the circular Wilson loop.

%Ark
For generic location of the dilaton operator one can check that the result is consistent with the general expectation (\ref{1.2}) (we have checked this explicitly for $n=5,6$.) For instance,
 for $n=6$ and the case of a regular polygon the result depends on three conformal ratios
 (since the polygon is regular only three cross-ratios are independent) and we obtain
\begin{equation}
F(\zeta_1,\zeta_2,\zeta_3)=-\frac{ \hat{c}_{dil}\  \lambda}{2\pi^4}
 \frac{   \zeta_1 \zeta_2 \zeta_3 ( \zeta_3-1)  + \zeta_3^2 - \zeta_2^3}
{\Big[\zeta_1 \zeta_2^2 \zeta_3^2 (\zeta_2-\zeta_3)^2 \Big(  \zeta_1  \zeta_3 ( \zeta_2-1)- \zeta_2^2   +\zeta_3\Big) \Big]^{1/3}}\ , \la{619}
\end{equation}
%F(\zeta_1,\zeta_2,\zeta_3)= \frac{\zeta_3^2(2+\zeta_1+\zeta_2+\zeta_1 \zeta_2)+\zeta_3(3+\zeta_1+\zeta_2+\zeta_1 %\zeta_2)-\zeta_2(3+3\zeta_2+\zeta_2^2)}{2\left((1+\zeta_1)(1+\zeta_2)(\zeta_2-\zeta_3)%(1+\zeta_3)^2(\zeta_3+\zeta_2(-1+\zeta_1-\zeta_2+\zeta_3+\zeta_1 \zeta_3)) \right)^{1/3}}
%\end{equation}
where the conformal ratios are defined  by
\begin{eqnarray}
&&\zeta_1= \frac{|x^{(1)}-x^{(3)}|^2|a-x^{(5)}|^2}{|x^{(1)}-x^{(5)}|^2|a-x^{(3)}|^2} \ ,  \ \ \ \ \ \ \ \ \ 
\zeta_2= \frac{|x^{(2)}-x^{(4)}|^2|a-x^{(6)}|^2}{|x^{(2)}-x^{(6)}|^2|a-x^{(4)}|^2}\ , \no \\
&&\zeta_3= \frac{|x^{(1)}-x^{(4)}|^2|a-x^{(3)}|^2|a-x^{(6)}|^2}{|x^{(3)}-x^{(6)}|^2|a-x^{(1)}|^2|a-x^{(4)}|^2}
\ . \la{zee}
\end{eqnarray}

%%%%%%%%%%%%%%%%%%%%%%%%%%%%%%%%%%%%%%%%%%%%%%%%%%%%%%%%%%%%%%%%%%%%%%%%%%%%%%%%%%%%%%%%%%%%%%%%%%%%%%%%%%%%%%%%%%%%%%%

\section{Concluding remarks}

In this  paper we considered the correlator  \rf{1.1} of a null $ n$-polygon  Wilson loop
with a local  operator, such as the  dilaton  ($\OO_{dil} \sim \tr F^2_{mn} +...$)or a chiral primary operator.
 Based on  symmetry considerations  we determined  its general form \rf{1.2}, expressing it in terms of
a function $F$  of $3n-11$ conformal ratios involving the position of the operator
and the positions of the  cusps.

 In the  first non-trivial case of $n=4$
 this function $F$   depends on just  one conformal ratio $\z$ (defined
 in \rf{ges}),  making the  corresponding correlator
\rf{1.1},\rf{1.15} one of the simplest non-trivial  observables
 that one would
like eventually  to compute exactly for all values of the `t Hooft  coupling $\l$.
The value of  $F$ determines, in particular,     the corresponding
 OPE coefficient \rf{1.166} in the expansion \rf{22}  of the Wilson loop in terms of
 local operators.

We have  found the leading terms in $F$    both at  strong coupling
 (using
semiclassical string theory)  and at weak  coupling (using  perturbative planar
gauge theory).
At leading order at strong coupling we find that
$F \sim {\sql } $  and has non-trivial dependence on $\z$
 \rf{2.17}  while at leading order in weak coupling $F \sim   {\l } $ and is constant
 \rf{6.12}.
In the case  of more general dilaton operator with non-zero R-charge $j$ (with  $\D=4+j$)
the strong-coupling expression for $F$ is given  by a hypergeometric function  \rf{2.33}.
Similar results were  found in the case of the chiral primary operator \rf{2.48},\rf{2.51}.

It would  be  important  to compute  subleading terms in the two respective expansions:
\bea \la{71}
&&F_{\l \gg 1} = { 1 \ov N} \big[ \sql   f_0 (\z)
 +  f_1(\z) + {1 \ov \sql} f_2( \z) + ... \big] \  , \\
&&F_{\l \ll 1} = { 1 \ov N} \big[  \l   h_0   +  \l^2 h_1(\z) + \l^3  h_2( \z) + ... \big] \  . \la{72}
\eea
Another open problem is the extension to the  case  of the $n >4$ cusped Wilson loop.
%A27
%So far we have managed to find  only numerical results for the OPE coefficient
%at strong coupling, see \rf{six}.

%Ark
Let us note that in the case of the dilaton operator integrating \rf{1.1} 
over the point $a$ we get the insertion of  the action and so the resulting correlator 
should be proportional to a  derivative over $\l$ of the logarithm of the null-polygon Wilson loop.
Thus, in particular,  the knowledge of $\lg W_n \rg $  at higher orders in $\l$ provides a constraint on 
integral of \rf{1.1}  at lower order order in $\l$; in general,  this is not, however, 
enough to determine the functions $h_n(\z)$ in \rf{72}.

Part of the original motivation for the present work  was to
shed  more light on the  relation \ci{ak} between a
correlator of null-separated  local operators  and the  square of
 corresponding   cusped    Wilson loop.  We conjectured
%AT
 %As was  noted in the Introduction, this suggests  a conjecture
a more general relation \rf{121}
% relating
connecting  correlators with one extra  operator at an arbitrary position
 to the  correlator \rf{1.1} we   considered in this paper.  It
 would  be interesting  to try to  verify  the relation  \rf{121}   for $n=4$
 at weak coupling.

There are several possible extensions of our present work.
One may  consider  the case when the local operator $\OO$
is not ``light''   at strong coupling   but is allowed  to
carry a large charge
(e.g.,  R-charge or angular momentum in $S^5$ so that $\D \sim \sql$).  As in  the circular loop case in \ci{za02},
then the semiclassical surface will need to be  modified to account for the
presence of the sources provided by the  vertex operator $\VV$ in the string path integral
(see also \ci{alt}).\foot{For example, one may consider  the operator inserted at
$|a|=\infty$. The resulting semiclassical world surface should then  have a topology of a disc with  a puncture, i.e. of a cylinder. For example,
in the case of a dilaton with  large charge $j$ we would  have
then a source provided by the vertex operator  $ \sim
a^{-2\D} \int du dv \ z^\D  e^{i j\varphi} U_{dil}$.
A naive  candidate for such  modified surface  is the  generalization
 \rf{2.36} of the regular 4-cusp  surface to the presence of $S^5$ angular
momentum density.  However, this surface has zero dilatation  charge
$z^{-2} ( z \del_u z + x_m \del_u x^m)$, while to support the source provided
by $z^\D$ one needs a surface with a dilatation charge  proportional  to $\D$
(cf.  \ci{za02}).
It remains an open question how  to find such a surface.}

One may   consider also  a correlator of a Wilson loop with
several ``light''
($\D \ll \sql$)
operators. At leading order in strong-coupling expansion  such a correlator
  should
 factorize like in the case of the correlators
two ``heavy'' ($\D \sim \sql$)  operators  and several ``light'' ones \ci{rt,bt2}, i.e.
 $ \lg  W_n  \OO(a_1) \OO(a_2) \rg \sim  \lg  W_n  \OO(a_1)\rg \lg  W_n  \OO(a_2)\rg $.
 This follows from the fact  that for $\sql \gg 1$ these   correlators
 are found, like in \rf{2.3},  by evaluating the  corresponding   vertex operators
 on the world surface  ending on the null polygon that defines $W_n$.\foot{Contribution
 of a more singular term  proportional to a power of  $|a_1 - a_2|^{-2}$
 is suppressed at large $\l$.
 Note also that  in general
 $ \lg  W_n  \OO(a_1) \OO(a_2) \rg$ should depend, say for $n=4$,
  on $3n-11 + 4 = 5$
 conformal ratios, this strong-coupling factorization implies
 that the leading term depends only on $ 2 \times (3n-11) = 2$  conformal ratios.}
 The study of such more general correlators may be of interest in
   trying to understand  better the relation \ci{ak} between the
correlator of null-separated  local operators  and the  square of
 corresponding   cusped  Wilson loop.
 % (e.g., one may hope to
 %find some  iterative relation that increases the number of cusps).

%%%%%%%%%%%%%%%%%%%%%%%%%%%%%%%%%%%%%%%%%%%%%%%%%%%%%%%%%%%%%%%%%%%%%%%%%%%%%%%%%%%%%%%%%%%%%%%%%%%%%%%%%%%%%%%%%%%%%%%%

\section*{Acknowledgments}

We are  grateful to G. Korchemsky,    R. Roiban
%AT
and E. Sokatchev  for    stimulating  discussions
% of related topics
and useful suggestions.
The work of E.I.B. is supported by an STFC fellowship.
Part of this work was completed while AAT was visiting TPGU  at Tomsk  and he
acknowledges  the support of  the  RMHE grant 2011-1.5-508-004-016.

%%%%%%%%%%%%%%%%%%%%%%%%%%%%%%%%%%%%%%%%%%%%%%%%%%%%%%%%%%%%%%%%%

\appendix
\section{On the general  structure of the correlator ${\cal C}(W_n, {a})$}

%%%%%%%%%%%%%%%%%%%%%%%%%%%%%%%%%%%%%%%%%%%%%%%%%%%%%%%%%%%%%%%%%%%%%%%%%%%%%%%%%%%%%%%%%%%%%%%%%%%%%%%%%%%%%%%%%%%%%%%%%%%

Here we  complete   the proof in section 2.1  that eq. \rf{1.2} 
gives the general expression for the correlator in \rf{1.1}.
As we discussed in section 2.1, the conformal invariance implies 
that  the numerator function $\FF$ in  \rf{e1}, i.e. in
%A27
\be
{\cal C}(W_n, a)=
\frac{\FF(a,x^{(i)})}{\prod_{k=1}^n
|a-x^{(k)}|^{{2\ov n}\D}}\,,
\label{e11}
\ee 
should have  dimension $\D$  and  should  transform under the inversions 
so that $\C$ in \rf{e1}  changes by factor of $|a|^{2 \D}$, i.e. $\FF$
should transform according to \rf{tra}. 
In general,  one may start with  $\FF$  
 written as a sum of several structures\foot{Note that this assumption is not a 
 restriction on generality
 as this sum may be infinite. Alternatively, starting, say, with $\FF$  written as a ratio 
 of similar sums one may repeat the argument below   by observing that  
 the numerator and denominator should have  definite dimensions, etc.}
\be
\FF= \sum_{I=1}^M  f_I(|x^{(ij)}|) \  F_I (\z) \,,
\label{e2}
\ee
where 
 $f_I$ are some functions of the distances
$|x^{(ij)}|=|x^{(i)}- x^{(j)}|$ between non-adjacent cusps
having  dimension  $\Delta$  and the
functions  $F_I(\zeta)$ depend only on conformal ratios $\z_k(a,x^{(i)})$.
%A27
Since all the  functions
$F_I$ are assumed to be independent, the structure of \rf{e11} 
implies   that under the inversions $f_I$ should transform as 
 \be 
 f'_I = \Big(|x^{(1)}|\ldots |x^{(n)}|\Big)^{-\frac{2
\Delta}{n}} f_I\,.
\label{e5}
\ee
Thus  all $f_I$'s  transform the same way  under the  conformal group.
 This means
that the ratio of any two of them is conformally invariant and,
hence, is a functions of conformal ratios. Therefore, we can
represent $\FF$  as follows
\be
\FF=  f_1  \Big(1+ \sum^M_{I=2}\frac{f_I}{f_1} F_I \Big) \equiv\ {f_1}\ F \,.
\label{e6}
\ee
Here the combinatioin in the brackets is conformally invariant and we 
denoted it as $F(\zeta)$. The  coefficient $f_1$ can be chosen  
so that  it has dimension $\Delta$ and transforms
as~\eqref{e5} under inversions:\foot{Note that we have a freedom in
determining $f_1$ since we can multiply it by an arbitrary
conformally invariant function.} 
\be
f_1 = \prod_{i< j-1}^n |x^{(i)}-x^{(j)}|^{\mu}\,, \ \ \ \ \ \ \ \ 
\ \m=\frac{2 }{n (n-3)}\Delta\,.
\label{e7}
\ee
Here the power $\mu$ is fixed  by taking into account  that 
  $f_1$  should   have dimension $\Delta$ (the number of diagonals of
the polygon is $\frac{n(n-3)}{2}$). 
Note that with this power $f_1$ also has the right transformation
property under the inversions.
We  thus arrive at the  general
expression \rf{1.2} for  ${\cal C}(W_n, a)$ in terms of a single function $F$
depending on $3n-11$ conformal ratios.

To get a better idea of the structure of \rf{1.2}  it is useful to look at specific examples.
In section 2.2 we discussed the case of $n=4$ cusps and here we will consider the $n=5$ case. 
In this  case there are 5  non-zero ``diagonals'' 
$|x^{(13)}|,|x^{(14)}|,|x^{(24)}|,|x^{(25)}|,|x^{(35)}|$ and $3n-11=4$ 
 non-trivial conformal ratios that may be chosen as (cf. \rf{ges}) 
\iffalse
\bea 
&&
\zeta_1= \frac{ |a - x^{(2)} |^2\ | a  - x^{(4)} |^2 \  | x^{(13)} |^2 }
  { |a- x^{(1)} |^2\ |  a- x^{(3)} |^2 \ | x^{(24)} |^2 } \ , \ \ \ \ 
  \zeta_2= \frac{ |a - x^{(2)} |^2\ | a  - x^{(5)} |^2 \  | x^{(13)} |^2 }
  { |a- x^{(1)} |^2\ |  a- x^{(3)} |^2 \ | x^{(25)} |^2 } \ ,\no \\
 && \zeta_3= \frac{ |a - x^{(2)} |^2\ | a  - x^{(5)} |^2 \  | x^{(14)} |^2 }
  { |a- x^{(1)} |^2\ |  a- x^{(4)} |^2 \ | x^{(25)} |^2 } \ , \ \ \ \ 
  \zeta_4= \frac{ |a - x^{(3)} |^2\ | a  - x^{(5)} |^2 \  | x^{(14)} |^2 }
  { |a- x^{(1)} |^2\ |  a- x^{(4)} |^2 \ | x^{(35)} |^2 } \ , \la{gees}
  \eea
  \fi
 \bea 
&&
\zeta_1= \frac{ |a - x^{(1)} |^2\  | x^{(24)} |^2 }
  { |a- x^{(2)} |^2\  | x^{(14)} |^2 } \ , \ \ \ \ 
  \zeta_2= \frac{ |a - x^{(2)} |^2 \  | x^{(35)} |^2 }
  { |a- x^{(3)} |^2 \ | x^{(25)} |^2 } \ ,\no \\
 && \zeta_3= \frac{ |a - x^{(3)} |^2\   | x^{(14)} |^2 }
  { |a- x^{(4)} |^2\  | x^{(13)} |^2 } \ , \ \ \ \ 
  \zeta_4= \frac{ |a - x^{(4)} |^2\   | x^{(25)} |^2 }
  { |a- x^{(5)} |^2\  | x^{(24)} |^2 } \ , \la{gees}
  \eea 
 so that  \rf{1.2} takes the following   explicit 
 form
\be
{\cal C}(W_5, a)= \frac{\big(|x^{(13)}||x^{(14)}||x^{(24)}||x^{(25)}||x^{(35)}|\big)
^{ {1 \ov 5} \D}}
{\big(|a- x^{(1)}| |a- x^{(2)}| |a- x^{(3)}||a- x^{(4)}| |a- x^{(5)}|\big)^{{2\ov 5}\D}}
\ F(\z_1, \z_2,\z_3,\z_{4})\,,
\label{1.26}
\ee
For example, in the  case of the  dilaton operator  with $\D=4$  
 the prefactor in \rf{1.26} contain  distances in power $1/5$; one may wonder how such 
 powers may appear in a perturbative  computation. 
To understand why there is no contradiction   let us consider a model expression that 
has all the required symmetries  and  yet contains 
only integer powers of distances  and then  show  that it  can be put into
 the general form \rf{1.26}.  Namely,   with $n=5$ and $\D=4$  the following expression 
 that mimicks the $n=4,\ \D=4$  one in \rf{1.15}  has the right scaling dimension and is
 covariant under the inversions:
 \bea 
&& \tilde {\cal C}(W_5, a)= 
\frac{  q_1\  |x^{(13)}|^2\  |x^{(24)}|^2 }  
{  |a -x^{(1)}|^2|a -x^{(2)}|^2 |a -x^{(3)}|^2 |a -x^{(4)}|^2}\no \\
&&+
\frac{ q_2 \ |x^{(53)}|^2\  |x^{(24)}|^2 }  
{  |a -x^{(5)}|^2|a -x^{(2)}|^2 |a -x^{(3)}|^2 |a -x^{(4)}|^2}+
\frac{ q_3 \ |x^{(13)}|^2\  |x^{(25)}|^2 }  
{  |a -x^{(1)}|^2|a -x^{(2)}|^2 |a -x^{(3)}|^2 |a -x^{(5)}|^2}\no \\ && +
\frac{ q_4 \ |x^{(14)}|^2\  |x^{(25)}|^2 }  
{  |a -x^{(1)}|^2|a -x^{(2)}|^2 |a -x^{(5)}|^2 |a -x^{(4)}|^2} +
 \frac{ q_5 \ |x^{(35)}|^2\  |x^{(14)}|^2 }  
{  |a -x^{(1)}|^2|a -x^{(5)}|^2 |a -x^{(3)}|^2 |a -x^{(4)}|^2}
\,,\no
\label{1.75}
\eea
where $q_i$ are conformal invariants (e.g., constants). 
It is straightforward to check that  each term here can be rewritten in the form of prefactor in 
\rf{1.26}  multiplied by an appropriate factor of the conformal ratios  \rf{gees}. 
The result is then given by \rf{1.26} with $F$  being a  sum of $F_i$'s multiplied by  powers
of $\z_k$. Explicitly, 
\be  F=    q_1 \ \big( \z_1 \z_2^2 \z_3^3  \z_4^4 \big)^{-1/5} + ...  \ee

\section{Some analytic results for even $n$ for  $\Delta \gg 1 $}

%%%%%%%%%%%%%%%%%%%%%%%%%%%%%%%%%%%%%%%%%%%%%%%%%%%%%%%%%%%%%%%%%%%%%%%%%%%%%%%%%%%%%%%%%%%%%%%%%%%%%%%%%%%%%%%%%%%%%%%%%

In this Appendix, we briefly describe some analytic results for regular polygons with even number of sides $n$ in the limit of large $\Delta$.
For larger and larger values of $\Delta$ the main contribution to the integral comes
from the region close to the origin. We can then solve the auxiliary
 linear problems     \rf{lin}
 perturbatively around $\rho=0$.
For instance, we have
\begin{equation}
\alpha= c_0+c_1\rho^2+c_2 \rho^4+ \dots\,, \ \ \ \ \ \ \ \ \ \ \ \rho= |\xi| \ ,
\label{B1}
\end{equation}
where
all the subsequent coefficients can be determined  in terms of $c_0$.
The coefficient $c_0$ is fixed by the boundary conditions at infinity and is
given by~\cite{McCoy}
\begin{equation}
e^{c_0}=3^{1/3}\frac{\Gamma[{2\ov 3}]}{\Gamma[{1\ov 3}]}\,.
\label{B2}
\end{equation}
Using this expansion, we can integrate the flat connections
along the radial direction, for an arbitrary angle. This discussion is pretty
general and applies to any $n$ (for the class of regular polygons which can be embedded into $AdS_3$). For definiteness, let us focus on  the case of the hexagon, $n=6$.
In this case we obtain
\begin{equation}
Y_{-1}=1+2 e^{2c_0}\rho^2+\frac{2}{3}e^{2c_0}(2c_1+e^{2c_0})\r^4  +...\  .
\label{B3}
\end{equation}
%{\bf is that right -- $\rho^4$  was  missed in the last term ? ?}
%
After performing the integrals we get
for the coefficient  in \rf{444}
% (ignoring the obvious factor $c_{dil}/|{a}|^8$)
%
\begin{equation}
I_{6,\Delta} = \frac{4 \pi}{\Delta}+\frac{4 \pi}{\Delta^2}+ \frac{12- 2e^{-6 c_0}}{3 \Delta^3}+...
\end{equation}
where $e^{c_0}$ is given by~\eqref{B2}.
%\begin{equation}
%e^{c_0}=3^{1/3}\frac{\Gamma[2/3]}{\Gamma[1/3]}
%\end{equation}
%This gives
%\begin{equation}
%I_{3,\Delta} = \frac{\pi}{2\Delta}+\frac{\pi}{2\Delta^2}+\frac{\pi}{12 \Delta^3}(6-\frac{\Gamma[1/3]^6}{9 \Gamma[2/3]^6})+...
%\end{equation}
As a check, for $\Delta=4$ this expression reduces to $I_6 \approx 3.90536$
%$I \approx 0.48817$,
which is very close from the actual numerical value $I_6=3.90901$ in \rf{six}.

\def \bi {\bibitem}
%%%%%%%%%%%%%%%%%%%%%%%%%%%%%%%%%%%%%%%%%%%%%%%%%%%%%%%%%%%%%%%%%%%%%%%%%%%%%%%%%%%%%%%%%%%%%%%%%%%%%%%%%%%%%%%%%%%%%%%

%\end{document}

\end{document}